\documentclass[aps,prd,superscriptaddress,nofootinbib,eqsecnum,twocolumn]{revtex4-1}

\pdfoutput=1

\usepackage{graphicx}
\usepackage{bm}
\usepackage{dcolumn}
\usepackage{amssymb}
\usepackage{amsmath}
\usepackage{epsfig} 
\usepackage{subfigure}
\usepackage[english]{babel}
\usepackage[latin1]{inputenc}
\usepackage{eucal}
\usepackage{hyperref}
\usepackage{verbatim}
\usepackage{latexsym}
\usepackage{color,xcolor}
\usepackage{ulem}
\usepackage{soul} 

\hypersetup{colorlinks=true,linkcolor=blue,filecolor=magenta,urlcolor=cyan,citecolor=orange}

\begin{document}

\title{Deconfinement and chiral phase transitions in quark matter with chiral imbalance}

\author{Francisco X. Azeredo} 
\affiliation{Departamento de F\'{\i}sica, Universidade Federal de Santa Maria, 97105-900, Santa Maria, RS, Brazil}

\author{Dyana C. Duarte}
\affiliation{Departamento de F\'{\i}sica, Universidade Federal de Santa Maria, 
97105-900, Santa Maria, RS, Brazil}

\author{Ricardo L. S. Farias} 
\affiliation{Departamento de F\'{\i}sica, Universidade Federal de Santa Maria, 
97105-900, Santa Maria, RS, Brazil} 

\author{Gast\~ao  Krein} 
\affiliation{Instituto de F\'{\i}sica Te\'orica, Universidade Estadual
  Paulista,  01140-070 S\~ao Paulo, SP, Brazil}

\author{Rudnei O.  Ramos}
\affiliation{Departamento de F\'{\i}sica Te\'orica, Universidade do
  Estado do Rio de Janeiro, 20550-013 Rio de Janeiro, RJ, Brazil}  

\begin{abstract}

We study the thermodynamics of the Polyakov-Nambu-Jona-Lasinio model considering the effects of an effective chiral chemical potential. We offer a new parametrization of the Polyakov-loop potential depending on temperature and the chiral chemical potential which, when used together with a
proper regularization scheme of vacuum contributions, predicts results consistent with those from lattice simulations.

\end{abstract}

\maketitle

\section{Introduction}
\label{intro}

There is significant interest in the literature in systems exhibiting nonzero chirality.  Such systems can exhibit a whole range of unusual and unconventional physical phenomena, such as the chiral magnetic effect (CME)~\cite{Fukushima:2008xe,Vilenkin:1980fu}, wherein a magnetic field applied to chirality-imbalanced matter induces a vector current; the chiral separation effect (CSE)~\cite{Son:2004tq,Metlitski:2005pr}, which describes the
induction of an axial current by a magnetic field in quark or ordinary
baryonic matter; and the chiral vortical effect (CVE)~
\cite{Son:2009tf,Vilenkin:1979ui,Banerjee:2008th,Landsteiner:2011cp},
where a magnetic field can prompt an anomalous current through the
CME; while a vortex in a relativistic fluid can also induce a current
via the chiral vortical effect. Another interesting outcome is the
Kondo effect, which can be driven by a chirality
imbalance~\cite{Suenaga:2019jqu}, among
others~\cite{Kharzeev:2010gd,Chernodub:2015gxa,Yamamoto:2015ria,Rajagopal:2015roa,Sadofyev:2015hxa}.
As a consequence of those diverse types of phenomena that are
predicted, the study of the effects of having a chiral medium find
applications in diverse physical systems of interest, such as, for
example, in the studies related to heavy-ion
collisions~\cite{Kharzeev:2024zzm}, in Weyl~\cite{Song:2016ufw} and
Dirac semimetals~\cite{Braguta:2019pxt}, in applications to understand
processes in the early Universe~\cite{Kamada:2022nyt} and compact
objects~\cite{Charbonneau:2009ax,Ohnishi:2014uea}, just to cite a few
(for a recent review discussing different applications, see, e.g.,
Ref.~\cite{Yang:2020ykt}). 

The effects of a chiral imbalance in a medium can be implemented in
the grand canonical ensemble by introducing a chiral chemical
potential $\mu_5$.  Chiral asymmetry effects on quark matter and
applications for the quantum chromodynamics (QCD) phase diagram
dualities were developed in
Refs.~\cite{Khunjua:2021oxf,Khunjua:2018jmn,Khunjua:2020vrp,Khunjua:2018sro}.
The influence of external electric and magnetic fields and $\mu_5$ on the QCD chiral phase transition was investigated
in~\cite{Ruggieri:2016xww}.  As far applications to QCD are concerned,
one very interesting aspect is related to the fact that QCD at finite
chiral chemical potential is amenable to lattice simulations, since it
is free from the sign problem, contrary to what happens, for example,
for the case of a finite baryon chemical potential. In the lattice
studies that implemented the effect of a chiral chemical
potential~\cite{Braguta:2015owi,Braguta:2015zta} it was predicted in
particular that the critical temperature ($T_c$) for chiral symmetry
restoration {\it increases} with $\mu_5$.  
However, the predictions from the literature based on effective models like the
Nambu-Jona-Lasinio (NJL)
model~\cite{Ruggieri:2011xc,Fukushima:2010fe,Chao:2013qpa,Yu:2014sla,Yu:2015hym}
and the quark linear sigma
models~\cite{Ruggieri:2011xc,Chernodub:2011fr} have found exactly the
opposite behavior, i.e., those model studies have shown a critical
temperature which is a {\it decreasing} function of  $\mu_5$ instead.
It was argued in Ref.~\cite{Braguta:2016aov} that $\mu_5$ can favor
quark-antiquark pairing, increasing the quark condensate and,
consequently, requiring a higher temperature for the  chiral
symmetry restoration, which is consistent with the results predicted
in Refs.~\cite{Braguta:2015owi,Braguta:2015zta}.  An agreement between
the behavior for $T_c$ as a function of $\mu_5$ as predicted by
lattice simulations was obtained by universality arguments in the
large $N_c$ limit (where $N_c$ is the number of color degrees of
freedom) in Ref.~\cite{Hanada:2011jb}, in studies with
phenomenological quark-gluon interactions in the framework of the
Dyson-Schwinger equations
\cite{Wang:2015tia,Xu:2015vna,Shi:2020uyb,Tian:2015rla,Cui:2016zqp},
nonlocal NJL
models~\cite{Frasca:2016rsi,Ruggieri:2016ejz,Ruggieri:2020qtq},
through the use of a self-consistent mean field approximation in the
NJL
model~\cite{Yang:2020ykt,Liu:2020elq,Liu:2020fan,Yang:2019lyn,Pan:2016ecs,Lu:2016uwy},
using a nonstandard renormalization scheme in the quark linear sigma
model~\cite{Ruggieri:2016cbq}, and also in the context of chiral
perturbation
theory~\cite{Espriu:2020dge,GomezNicola:2023ghi,Andrianov:2019fwz}. 

In a previous work~\cite{Farias:2016let}, it was also addressed the
problem observed between the local NJL model results and those
obtained from the lattice when concerning the behavior of $T_c$ in
terms of $\mu_5$. The investigation carried out in Ref.~\cite{Farias:2016let} has revealed that the disagreement between
the lattice with the model results could be explained by the way
momentum integrals of vacuum quantities were treated within the NJL
model. More specifically, it was shown that the discrepancy between lattice and model results can be eliminated with a proper treatment of the integrands of the divergent momentum integrals. The methodology of properly treating divergent integrals in the NJL model was named in Ref.~\cite{Farias:2016let} medium separation scheme
(MSS). The use of the MSS allows for the effective separation of medium
contributions from divergent integrals: notably, the remaining
divergences are the same as the ones encountered in the vacuum of the
model, at $T = \mu_5 = 0$.  As a consequence, the findings of Ref.~\cite{Farias:2016let} have pointed to an increase of $T_c$ with
$\mu_5$ and, thus, exhibiting qualitative agreement with the
prevailing physical expectations from the lattice results and also
confirmed by other more recent and more involved methods. This
correspondence of the effect of $\mu_5$ on the critical temperature is
consistent with the understanding that $\mu_5$ serves as a catalyst
for dynamical chiral symmetry breaking (DCSB)~\cite{Braguta:2016aov},
thereby implying an anticipated increase in the critical temperature
as a function of $\mu_5$.

In this work, we extend the methodology introduced in Ref.~\cite{Farias:2016let} to explore the dynamics of the
confinement-deconfinement transition influenced by $\mu_5$ within the
Polyakov extended NJL model (PNJL). This investigation aligns with
predictions from lattice QCD~\cite{Braguta:2015owi}, where the
critical temperatures associated with the confinement-deconfinement transition and the chiral symmetry restoration increase with
$\mu_5$. Notably, to the best of our knowledge, there is no framework that simultaneously captures the behavior of both (pseudo) criticial temperatures in accordance with lattice predictions. As we  show in this work, the agreement of the results requires the application of the MSS procedure once again but also a new parametrization of the Polyakov-loop potential in order to include the effect of the chiral chemical potential  besides of the standard one given in terms of the temperature only. In this work, we offer  such an appropriate parametrization to reproduce the lattice results.

The remainder of this paper is organized as follows. In
Sec.~\ref{sec2} we discuss the PNJL model studied in this work along
also with the applied regularization scheme. In Sec.~\ref{sec3} we
present our numerical results. We compare our regularization scheme
with the one more commonly considered in the literature. We propose a new parametrization of the Polyakov loop, depending on $T$ {\em and} $\mu_5$,  which turns out to be more appropriate to describe the effects due to a chiral chemical potential. A detailed analysis of the phase diagram of the model is
presented in Sec.~\ref{sec4}. {}Finally, our conclusions and final
remarks are presented in Sec.~\ref{conclusions}. 

\section{PNJL model and equations}
\label{sec2}

To include the effects of a baryon density and chiral imbalance, one
may start with the partition function in the grand canonical ensemble,
\begin{align}
\mathcal{Z}&(T,\mu_B,\mu_5) =
\int[d\bar{\psi}][d\psi]\times\nonumber\\ &\exp\left[\int\limits_0^\beta
  d\tau\int d^3x\left(\mathcal{L}_{\text{PNJL}} +
  \bar{\psi}\mu\gamma_0\psi +
  \bar{\psi}\mu_5\gamma_0\gamma_5\psi\right)\right]\,,
\nonumber\\
\label{ZTmu}
\end{align}
where $\mu = \text{diag}(\mu_u,\mu_d)$ is the quark chemical
potential, related to the baryon chemical potential as $\mu =
\mu_B/N_c$ in the isospin symmetric limit and $\mu_5$ is a pseudo
chemical potential related to the imbalance between right- and
left-handed quarks.  The Lagrangian density for the PNJL model with
$SU(2)$ symmetry is given by~\cite{Ratti:2005jh}
\begin{eqnarray}
 \mathcal{L}_{\text{PNJL}} & = & \bar{\psi}\left(i\gamma_{\mu}D^{\mu}
 - m_c\right)\psi + G\left[\left(\bar{\psi}\psi\right)^2 +
   \left(\bar{\psi}i\gamma_5\vec{\tau}\psi\right)^2\right]
 \nonumber\\ &
 & -~\mathcal{U}\left(\Phi[A],\Phi^{\dagger}[A],T\right),
\label{lagr}
\end{eqnarray}
where, in the isospin-symmetric limit, the current quark masses $m_c =
\text{diag}(m_u,m_d)$ have the same value, as will be discussed
below. In Eq.~(\ref{lagr}), $\mathcal{U}$ is the Polyakov loop
potential, $\psi = (\psi_u,\psi_d)^T$ is the quark field, $D^{\mu} =
\partial^{\mu} - iA^{\mu}$, with  $A^{\mu} = \delta_{\mu,0}A^0$. The
gauge coupling $g$ is absorbed in the definition of $A_{\mu}(x) =
g\mathcal{A}_a^{\mu}(x)\lambda_a/2$, where $\mathcal{A}_a^{\mu}(x)$ is
the $SU(3)$ gauge field and  and $\lambda_a/2$ are the Gell-Mann
matrices~\cite{Ratti:2005jh}. In this sense,   the scalar and
pseudoscalar effective coupling $G$ introduces the local (four-point)
interactions for the quark fields.  The order parameter for the
deconfinement phase transition in the pure gauge sector, $\Phi$,  and
its charge conjugate, $\Phi^{\dagger}$, are written in terms of the
traced Polyakov line with periodic boundary conditions,
\begin{equation}
\Phi = \frac{\mathrm{Tr}_c L}{N_c};~~\Phi^{\dagger} =
\frac{\mathrm{Tr}_c L^\dagger}{N_c},
\end{equation}
where 
\begin{eqnarray}
L(x) = \mathcal{P}\left\{\exp\left[\int\limits_0^\beta d\tau
  A_4(\vec{x},t)\right]\right\},
\end{eqnarray}
with $\beta = 1/T$ is the inverse of the temperature and $A_4 = iA_0$,
and the symbol $\mathcal{P}$ denotes the time-ordering in the
imaginary time $\tau$.  The temperature-dependent effective potential
$\mathcal{U}$ describes the phase transition characterized by the
spontaneous breaking of $Z_3$ symmetry: with the increasing of the
temperature it develops a second minimum at $\Phi\neq 0$, which
becomes the global minimum at a critical temperature $T_0$. At large
temperatures, $\Phi\to 1$, while at low temperatures $\Phi\to
0$. There are different parametrizations for the potential
$\mathcal{U}(\Phi,\Phi^{\dagger},T)$ (for a recent review, see,
e.g. Ref.~\cite{Fukushima:2017csk}). In this work we adopt the
polynomial function~\cite{Ratti:2005jh,Costa:2009ae}
\begin{eqnarray}
  \mathcal{U}(\Phi,\Phi^{\dagger},T)&=&T^{4}\bigg[-\frac{b_{2}(T)}{2}
    \Phi\Phi^{\dagger}-\frac{b_{3}}{6}(\Phi^{3}
  + {\Phi^{\dagger}}^{3})\nonumber\\ &&+\frac{b_{4}}{4}(\Phi
  \Phi^{\dagger})^{2}\bigg],
\label{Uphi}
\end{eqnarray}
with
\begin{equation}
  b_{2}(T)=a_{0}+a_{1}\left(\frac{T_{0}}{T}\right)+
  a_{2}\left(\frac{T_{0}}{T}\right)^{2}+
  a_{3}\left(\frac{T_{0}}{T}\right)^{3},
\end{equation}
with $T_{0}=270$ MeV and the constants $a_0,\ldots,a_3, b_3, b_4$ are
chosen such as to fit the pure gauge results in lattice QCD and they
are given in Table~\ref{Tab1}. It has been argued in the literature
that in the presence of light dynamical quarks one needs to rescale
the value  of $T_0$ to 210 MeV and 190 MeV for two or three flavor
cases, respectively, with an uncertainty of about 30
MeV.~\cite{Carlomagno:2021gcy,Schaefer:2009ui}. However, since we are
interested in reproducing the lattice QCD results in the presence of a
chiral imbalance~\cite{Braguta:2015owi}, in this work we keep the
original value of $T_0 = 270$ MeV.

\begin{table}[ht!]
\caption{\label{Tab1} Parameters for PNJL when using the polynomial
  parametrization.}
\begin{center}
\begin{tabular}{cccccc}
\hline 
\hspace*{.4cm}$a_{0}$\hspace*{.4cm}
& \hspace*{.4cm}$a_{1}$\hspace*{.4cm}
& \hspace*{.4cm}$a_{2}$\hspace*{.4cm}
& \hspace*{.4cm}$a_{3}$\hspace*{.4cm}
& \hspace*{.4cm}$b_{3}$\hspace*{.4cm}
& \hspace*{.4cm}$b_{4}$\hspace*{.4cm}\\  \hline  6.75 & -1.95 & 2.625
& -7.44  & 0.75 & 7.5 
\\\hline 
\end{tabular}
\end{center}
\end{table}

{}For the NJL model parameters, we follow~\cite{Costa:2009ae} and
adopt $\Lambda = $ 651 MeV, $G = $ 5.04 GeV$^{-2}$, $m_c = $ 5.5 MeV,
which reproduce the empirical values of the pion decay constant, $f_{\pi} = $ 92.3 MeV, the pion mass, $m_{\pi} = $ 139.3 MeV, and
the quark condensate, $-\left|\bar q q\right|^{1/3} = $ 251 MeV.  The final expression for the effective potential is given by 
\begin{align}
\Omega&(M,\Phi,\Phi^{\dagger},T,\mu,\mu_5) =
\mathcal{U}\left(\Phi,\Phi^{\dagger},T\right) +
\frac{\left(M-m_{c}\right)^{2}}{4G}\nonumber\\ &+ \Omega_V -
\frac{N_{f}}{\beta}\sum_{s=\pm1}
\int\frac{d^{3}p}{\left(2\pi\right)^{3}}
\log\left(F_{s}^{+}F_{s}^{-}\right) - C,
 \label{V1pnjl}
\end{align}
where $C$ is a constant added to ensure that the pressure vanishes in the vacuum, and
the functions $F_{s}^{+}$ and $F_{s}^{-}$ are
given by
\begin{eqnarray}
F_{s}^{+} & = & 1+3\Phi^{\dagger}
e^{-\beta\left(\omega_{s}+\mu\right)} +3\Phi
e^{-2\beta\left(\omega_{s}+\mu\right)}\nonumber\\ & + &
e^{-3\beta\left(\omega_{s}+\mu\right)}\,,\nonumber\\ F_{s}^{-} & = &
1+3\Phi e^{-\beta\left(\omega_{s}-\mu\right)} +3\Phi^{\dagger}
e^{-2\beta\left(\omega_{s}-\mu\right)}\nonumber\\ & + &
e^{-3\beta\left(\omega_{s}-\mu\right)}.
\end{eqnarray}
with $\omega_{s}= \sqrt{\left(|\mathbf{p}|+s\mu_{5}\right)^{2}+M^{2}}$ being the  $\mu_5$-modified dispersion relation.

Minimizing the thermodynamic potential in Eq.~(\ref{V1pnjl}) with respect to $M$, $\Phi$, and $\Phi^\dag$,  we obtain the following gap equations
\begin{equation}
\frac{\partial\Omega}{\partial M} 
= \frac{\partial\Omega}{\partial\Phi}
= \frac{\partial\Omega}{\partial\Phi^{\dagger}}= 0,
\label{gaps}
\end{equation}
where $\Omega = \Omega(M,\Phi,\Phi^{\dagger},T,\mu,\mu_5)$. These gap equations are required in the following when we discuss the different regularization schemes.
\subsection{Regularization}

The vast majority of studies using the NJL and the PNJL models
make use of a three-dimensional sharp cutoff $\Lambda$ to regularize
the divergent integrals appearing in $\Omega_V$. Since both models are
nonrenormalizable, $\Lambda$ becomes a scale for numerical results and a model parameter that must be determined together with the current quark mass $m_c$ and the coupling constant $G$. 

In this work, we compare the results obtained by the traditional
three-dimensional cutoff regularization scheme (TRS), and the MSS that have been frequently used in the
literature to describe the QCD phase structure in different
contexts~\cite{Carlomagno:2021gcy,Schaefer:2009ui,Carlomagno:2021gcy,Braguta:2015owi,Lopes:2021tro}.
The MSS regularization was first proposed in Ref.~\cite{Farias:2005cr}, and was first applied to a problem involving a chiral imbalance
in Ref.~\cite{Farias:2016let}. In the following, we summarize the implementation of this regularization scheme. We start from the mass gap equation at $T
= \mu = 0$:
\begin{eqnarray}
\frac{M-m_c}{2G} - M N_c N_f I_M = 0,
\end{eqnarray}
with
\begin{eqnarray}
I_M = \sum_{s=\pm 1}\int\frac{d^3p}{(2\pi)^3}\frac{1}{\omega_s}. 
\label{IM}
\end{eqnarray}
Since $M$ is an implicit function of $T, \mu$ and $\mu_5$, $I_M$ mixes
a vacuum part, including the divergences of the theory, and medium
contributions, that are finite and should not be regularized. The
incorrect regularization of medium contributions leads to different
nonphysical results as discussed in Ref.~\cite{Farias:2016let}. Equation~(\ref{IM}) can be rewritten as
\begin{equation}
I_M = \frac{1}{\pi}\int\limits_{-\infty}^\infty dx\sum_{s=\pm
  1}\int\frac{d^3p}{(2\pi)^3}\frac{1}{x^2 + \omega_s^2}.
\label{IM2}
\end{equation}
By iterating the identity
\begin{align}
\frac{1}{x^2 + \omega_s^2} &= \frac{1}{x^2 + p^2 + M_0^2}\nonumber\\  &
+ \frac{p^2 + M_0^2 - \omega_s^2}{(x^2 + p^2 + M_0^2)(x^2 +
  \omega_s^2)},
\end{align}
and performing some straightforward algebraic manipulations
(see~\cite{Farias:2016let} and~\cite{Duarte:2018kfd} for more explicit
details of the MSS implementation) one can express $I_M$ as
\begin{align}
I_M &= 2I_{\mathrm{quad}}(M_0) - (M^2 - M_0^2 -
2\mu_5^2)I_{\mathrm{log}}(M_0)\nonumber\\ &+ \left[\frac{3(M_0^2 - M^2
    - \mu_5^2)^2}{4}\right]I_1 + 2I_2 ,
\label{IMnew}
\end{align}
with the definitions
\begin{align}
&I_{\mathrm{quad}}(M_0) = \int\limits_0^{\Lambda}
  \frac{dp}{2\pi^2}\frac{p^2}{\sqrt{p^2 +
      M_0^2}},\\ &I_{\mathrm{log}}(M_0) = \int\limits_0^{\Lambda}
  \frac{dp}{2\pi^2}\frac{p^2}{\left(p^2 + M_0^2\right)^{3/2}},\\ &I_1
  = \int\limits_0^{\infty} \frac{dp}{2\pi^2}\frac{p^2}{\left(p^2 +
    M_0^2\right)^{5/2}},\\ &I_2  = \frac{15}{32}\sum_{s=\pm
    1}\int\frac{d^3 p}{(2\pi)^3}\int\limits_0^1
  dt(1-t)^2\nonumber\\ &\times \frac{(M_0^2 - M^2 - \mu_5^2 -
    2s\omega_p\mu_5)^3}{\left[(2s\omega_p\mu_5 - M_0^2 + M^2 +
      \mu_5^2)t + p^2 + M_0^2\right]^{7/2}} ,
\end{align}
and $\omega_p = \sqrt{p^2 + M^2}$. Note that $M_0$ is the effective
quark mass in the vacuum, i.e., it is the effective mass evaluated at
$T = \mu = \mu_5 = 0$, and it serves as a scale parameter for the
MSS. It can be determined from the model parametrization, given the
mass relation with the chiral condensate, $M = m_c - 4G\langle\bar q
q\rangle$. 

Note that in Eq.~(\ref{IMnew}) the medium contributions are completely
removed from the divergent integrals, and only
$I_{\mathrm{quad}}(M_0)$ and $I_{\mathrm{log}}(M_0)$, which are
explicit functions of $M_0$ are regularized.  Both $I_1$ and $I_2$ are
ultraviolet finite integrals and, therefore, they can both be
performed by extending the momentum integration up to infinity.

The thermodynamic potential is given by Eq.~\eqref{V1pnjl}. The only
different contribution for TRS and MSS is the term $\Omega_V$,
\begin{align}
\Omega_V^{\mathrm{TRS}} &= -
N_{c}N_{f}\sum_{s=\pm1}\int\limits_0^\Lambda\frac{dp~p^2}{2\pi^2}
\bigg[\omega_s - \omega_0\bigg],
\label{Vtrs}
\\ \Omega_V^{\mathrm{MSS}} &= -2N_{c}N_{f}\left\{\frac{M^2 -
  M_0^2}{2}I_{\mathrm{quad}}(M_0)\right.\nonumber\\ & \left.+
\left[\mu_5^2 M^2 + \frac{(M_0^2-
    M^2)^2}{4}\right]\frac{I_{\mathrm{log}}(M_0)}{2}  + I_{\mathrm
  {fin}}\right\},
 \label{Vmss}
\end{align}
with $\omega_0 = \sqrt{p^2 + M_0^2}$ along also with the definition
\begin{align}
I_{\mathrm{fin}} &=
\int\limits_0^\infty\frac{dp~p^2}{2\pi^2}\bigg[\frac{(M^2 - M_0^2)^2
    - 4M^2\mu_5^2}{8\omega_0^3} - \frac{M^2 -
    M_0^2}{2\omega_0}\nonumber\\ &  - \omega_0 +
  \frac{1}{2}\sum_{s=\pm1}\omega_s \bigg].
\label{Ifin}
\end{align}
The subtracted term $\omega_0$ in Eq.~(\ref{Ifin}) is a part of the
constant $C$ in Eq.~\eqref{V1pnjl}, it is required to cancel the divergences, allowing for the possibility of writing a finite expression for the thermodynamic potential in the MSS. Although the identification of $C$ is not trivial in the MSS, it is clear that in the TRS we have
\begin{equation}
C = \frac{(M_0 - m_c)^2}{4G}
-2N_{c}N_{f}\int\limits_0^\Lambda\frac{dp~p^2}{2\pi^2}\omega_0.
\end{equation}

An alternative version of the MSS thermodynamic potential may be
obtained from the mass gap equation,  which after performing all the
finite integrations analytically, we obtain
\begin{align}
&\frac{M-m_c}{2GMN_c N_f} = 2I_{\mathrm{quad}}(M_0) - (M^2 - M_0^2 -
  2\mu_5^2)I_{\mathrm{log}}(M_0)\nonumber\\  &- \frac{2\mu_5^2 + M^2 -
    M_0^2}{8\pi^2} + \frac{M^2 -
    2\mu_5^2}{8\pi^2}\ln{\left(\frac{M^2}{M_0^2}\right)}.
\label{gapeq}
\end{align}
Integrating Eq.~(\ref{gapeq}) with respect to $M$, we obtain 
 \begin{align}
\Omega_V^{\mathrm{MSS}} & =
-2N_{c}N_{f}\biggl\{\frac{M^2}{8\pi^2}\left[\frac{M^2}{4} -
  \mu_5^2\right]\ln\left(\frac{M^2}{M_0^2}\right)\nonumber\\ &-
\frac{3M^4}{64\pi^2} + \frac{M^2M_0^2}{16\pi^2} +
\frac{M^2}{2}I_{\mathrm{quad}}(M_0) \nonumber\\ &  + \left[\mu_5^2 M^2
  - \frac{M^4}{4} +
  \frac{M^2M_0^2}{2}\right]\frac{I_{\mathrm{log}}(M_0)}{2}\biggl\},
  \label{Vmssv2}
\end{align}
hence obtaining the result presented in Ref.~\cite{Farias:2016let}.

\begin{center}
\begin{figure}[!htpb]
{\includegraphics[width=8.2cm]{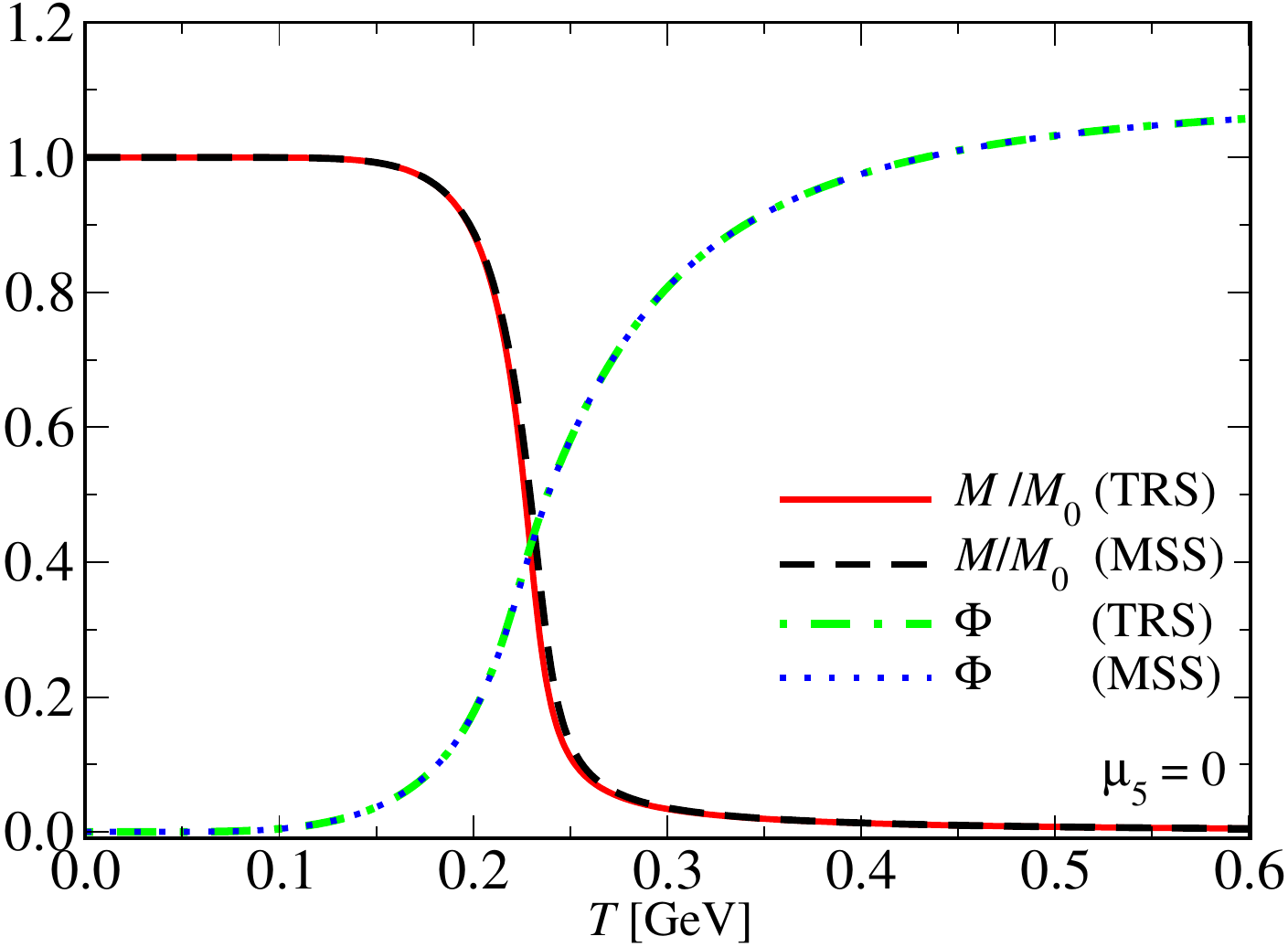}}
 \caption{Normalized quark mass $M/M_0$ and Polyakov loop $\Phi$ as a
   function of the temperature at $\mu = \mu_5 = 0$, for both
   regularization schemes.}
 \label{Fig1}
 \end{figure}
\end{center}

\begin{figure*}[t]
\begin{center}
\subfigure[]{\includegraphics[scale=0.33]{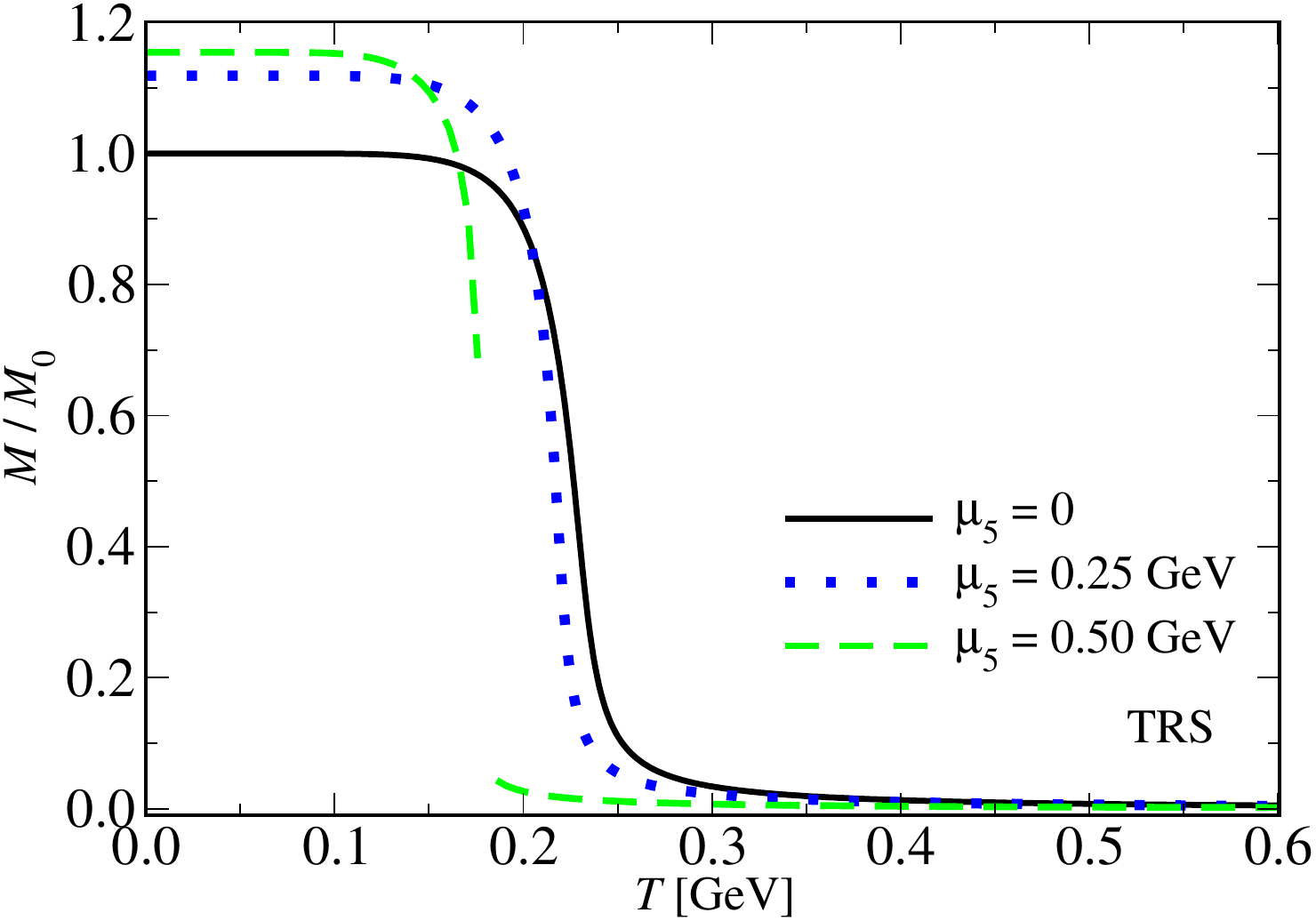}}
\subfigure[]{\includegraphics[scale=0.33]{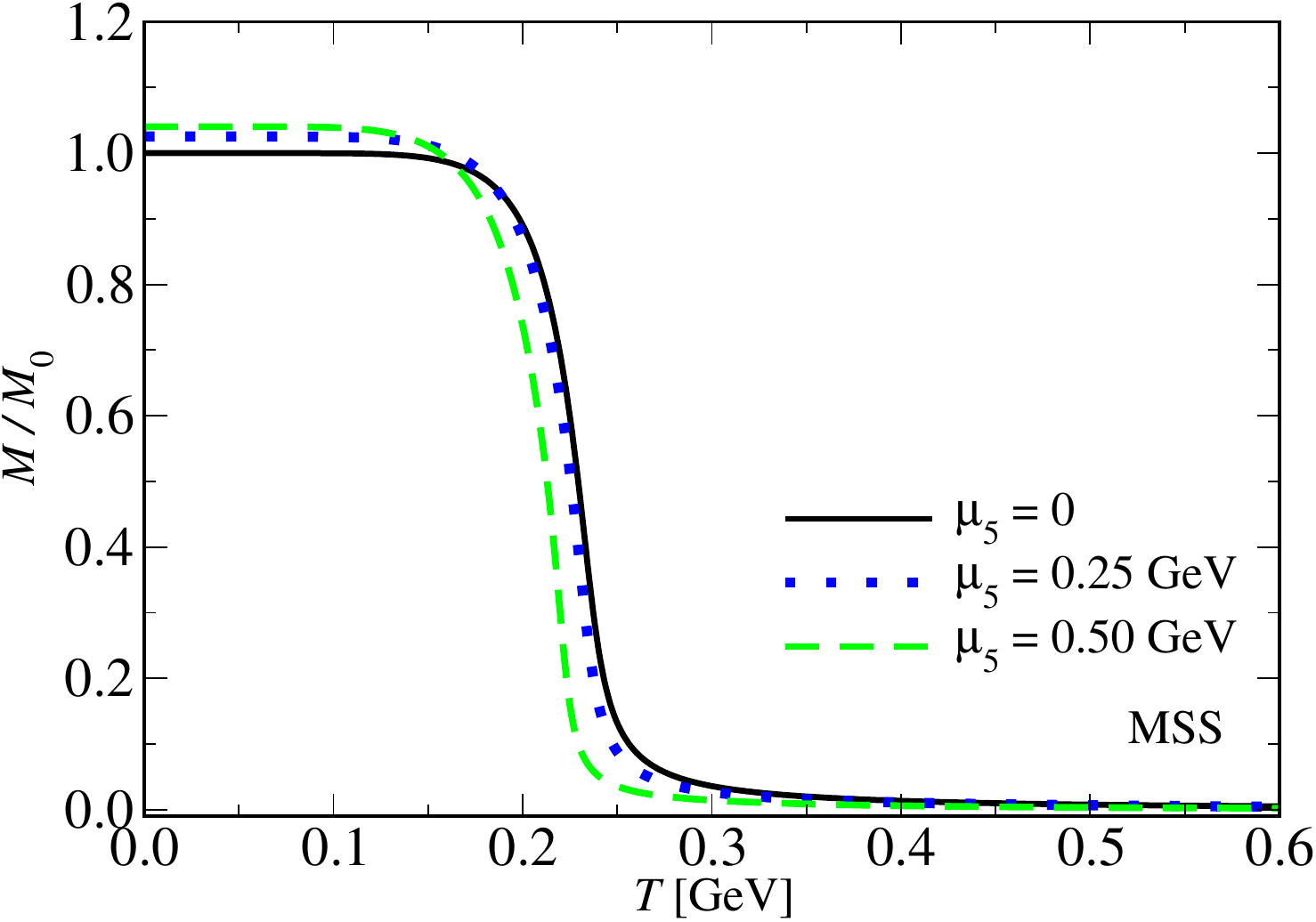}}\\
\subfigure[]{\includegraphics[scale=0.33]{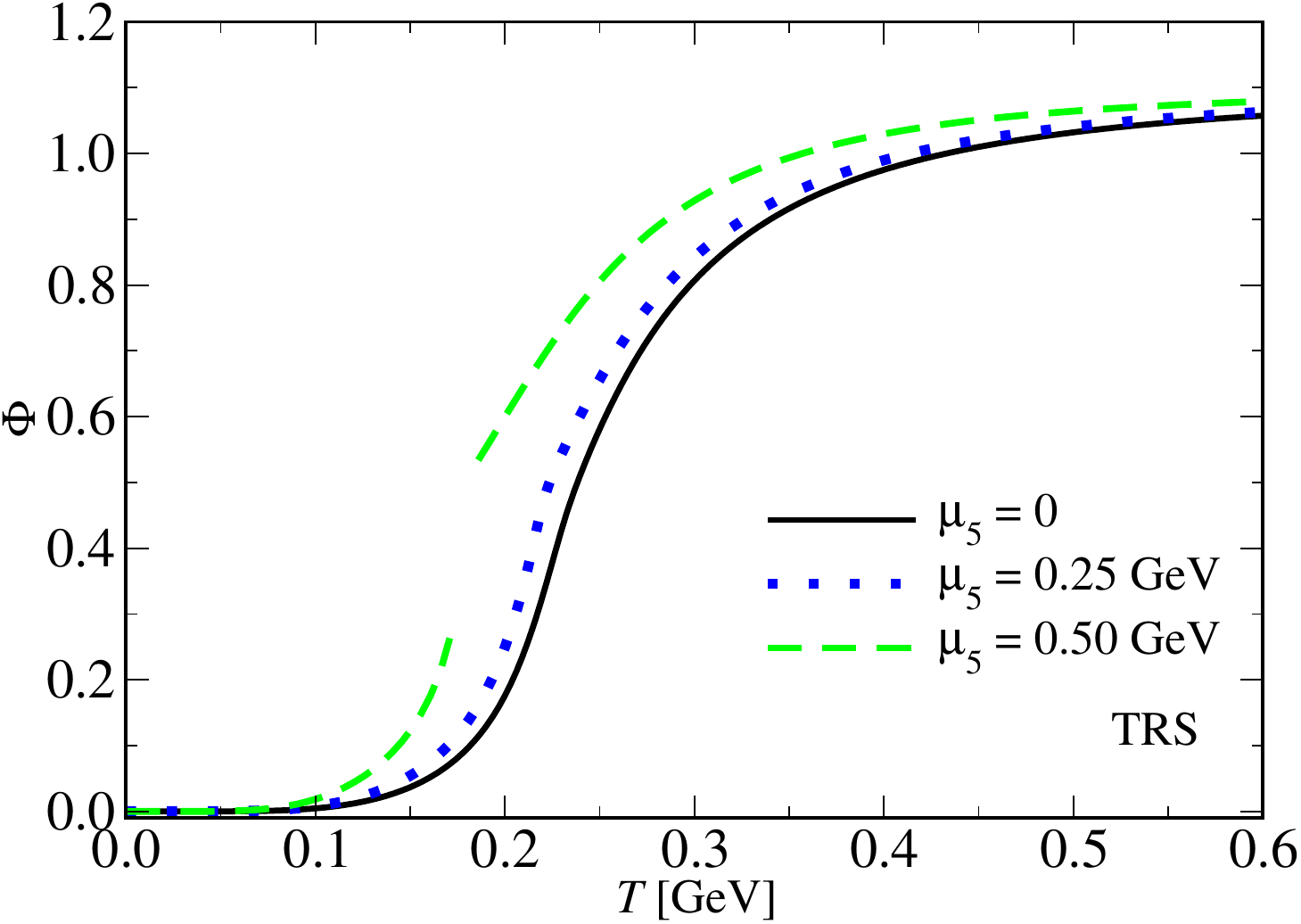}}\hspace{0.3cm}
\subfigure[]{\includegraphics[scale=0.33]{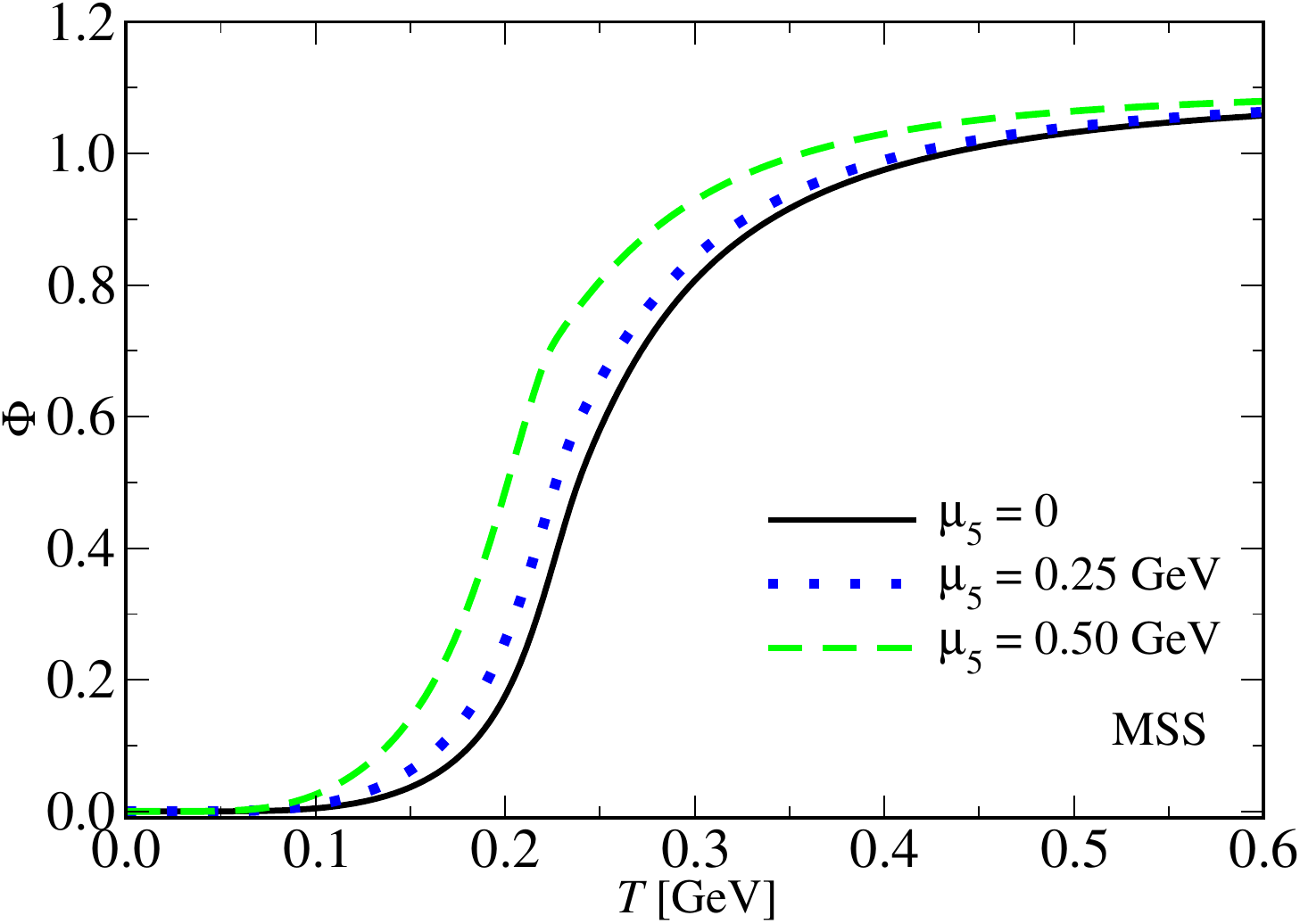}}
  \caption{Normalized quark mass $M / M_{0}$  and Polyakov loop $\Phi$
    in the TRS case [panels (a) and (c)] and in the MSS case [panels
      (b) and (d)] as functions of the temperature and for different
    values of $\mu_5$.}
 \label{Fig2}
 \end{center}
 \end{figure*}

As an illustration of the differences obtained for physical quantities
when evaluating them in the TRS and MSS methods, let us first show the
results for $M$ and $\Phi$, which are the solutions of
Eqs.~\eqref{gaps}. These results are  shown in {}Fig.~\ref{Fig1} for
$\mu = \mu_5 = 0$ and in {}Fig.~\ref{Fig2} for different values of
$\mu_5$.  It is worth mentioning that in the $\mu = 0$ case, we obtain
numerically $\Phi = \Phi^{\dagger}$, as previously discussed in the
literature~\cite{Ratti:2005jh}. Although the differences between TRS and
MSS are almost imperceptible in {}Fig.~\ref{Fig1} (even at $\mu_5=0$,
the MSS method still makes a medium separation effect due to the
temperature dependence of the effective mass $M$),
they get notably more pronounced in  {}Fig.~\ref{Fig2}. In
particular, both $M$ and $\Phi$ display a first-order phase transition for $\mu_5 \gtrsim 0.44$ GeV in the TRS case, while it is absent in the MSS method.

To motivate and emphasize the importance of the correct separation of medium contributions in the presence of $\mu_5$, we compare the results predicted by both schemes for the chiral density $\langle n_5\rangle$. {}From chiral perturbation
theory (ChPT), the QCD partition function in the presence of $\mu_5$ is modified to~\cite{Astrakhantsev:2019wnp}
\begin{equation}
Z(\mu_5) = Z_{\text QCD} \exp\left[\beta V N_f
  f_{\pi}^2\mu_5^2\right],
\end{equation}
where $Z_{\text QCD}$ is the finite-temperature QCD partition function. From this modified partition function one obtains for average of $\langle n_5\rangle$:
\begin{equation}
\langle n_5\rangle = \frac{1}{\beta V}\frac{\partial
  \log[Z(\mu_5)]}{\partial \mu_5} = 2N_f f_{\pi}^2\mu_5\,.
\label{n5}
\end{equation}
from which one concludes that $\langle n_5 \rangle$ is a linearly increasing function of $\mu_5$
with with slope $4f_{\pi}^2$ in the two-flavor case.  Within NJL model, the  chiral
density can be evaluated from Eq.~\eqref{V1pnjl} as
\begin{eqnarray}
\langle n_5\rangle = -\frac{\partial\Omega}{\partial\mu_5}\label{n5full}\,,
 \end{eqnarray}
which leads to:
\begin{equation}
\langle n_5\rangle^{\mathrm{TRS}} =
2N_{c}\sum_{s=\pm1}\int\limits_0^\Lambda\frac{dp~p^2}{2\pi^2}
\frac{s(|\mathbf{p}|
  + s\mu_5)}{\sqrt{(|\mathbf{p}| + s\mu_5)^2 + M^2}}.
\label{n5TRS}
\end{equation}
{}From this result, it is impossible to establish any connection
with the ChPT prediction since the $\mu_5$ effect is being regularized
together with the logarithmic divergence in the momentum
integral. {}For the MSS, on the other hand, the derivative
of~Eq.~\eqref{Vmssv2} results in
\begin{eqnarray}
\langle n_5\rangle^{\mathrm{MSS}} = 4N_c\left[M^2 I_{\mathrm{log}}(M_0)-
  \frac{M^2}{4\pi^2}\ln\left(\frac{M^2}{M_0^2}\right)\right]\mu_5\,,\nonumber\\
\label{n5mss1}
\end{eqnarray}
which is a linear function of $\mu_5$ and agrees with the linear behavior predicted by ChPT. Moreover, the slope is precisely the one predicted by ChPT, $4f_{\pi}^2$, as we show next.  To this end, we follow the works related to the origin of the MSS, referred to as the {\it
Implicit Regularization Scheme}~\cite{Battistel:1998tj,Farias:2005cr}. Specifically, the authors of those references showed that
\begin{eqnarray}
i \tilde{I}_{\text{log}}(M^2) = -\frac{f_{\pi}^2}{12M^2}\,.
\label{fpiIlog}
\end{eqnarray}
In this equation, $\tilde{I}_{\text{log}}(M^2)$ is defined in
Minkowski space as 
\begin{equation}
\tilde{I}_{\text{log}}(M^2) =
\int_{\Lambda}\frac{d^4k}{(2\pi)^4}\frac{1}{(k_0^2 - k^2 -  M^2)^2}\,,
\end{equation}
at an arbitrary mass scale $M$, which can still be a function of
$\mu_5$.\footnote{In Ref.~\cite{Farias:2005cr} the authors were
interested in the color superconducting gap $\Delta$ behavior, but
their approach may be used for the mass scale $M$ without loss of
generality.} We note that
$4i\tilde{I}_{\text{log}}(x^2) = I_{\text{log}}(x^2)$. This definition
is valid at any mass scale $M$, and can be expressed in terms of the vacuum quark mass, $M_0$, using the identity,
\begin{equation}
\tilde{I}_{\text{log}}(M^2) = \tilde{I}_{\text{log}}(M_0^2) -
\frac{i}{(4\pi)^2}\ln\left(\frac{M^2}{M_0^2}\right)\,.
\end{equation}
This allows us to rewrite Eq.~(\ref{n5mss1}) as
\begin{eqnarray}
\langle n_5\rangle^{\mathrm{MSS}}  & = & -16N_c
M^2i\tilde{I}_{\mathrm{log}}(M^2)\mu_5 {\color{blue}.}
\end{eqnarray}
With the use of $f_{\pi}^2$ from Eq.~\eqref{fpiIlog} and taking $N_c = 3$, leads to
\begin{equation}
\langle n_5\rangle^{\mathrm{MSS}} = 4f_{\pi}^2\mu_5\,,    
\end{equation}
which proves the ChPT result.

\begin{figure}[ht]
\includegraphics[scale=0.33]{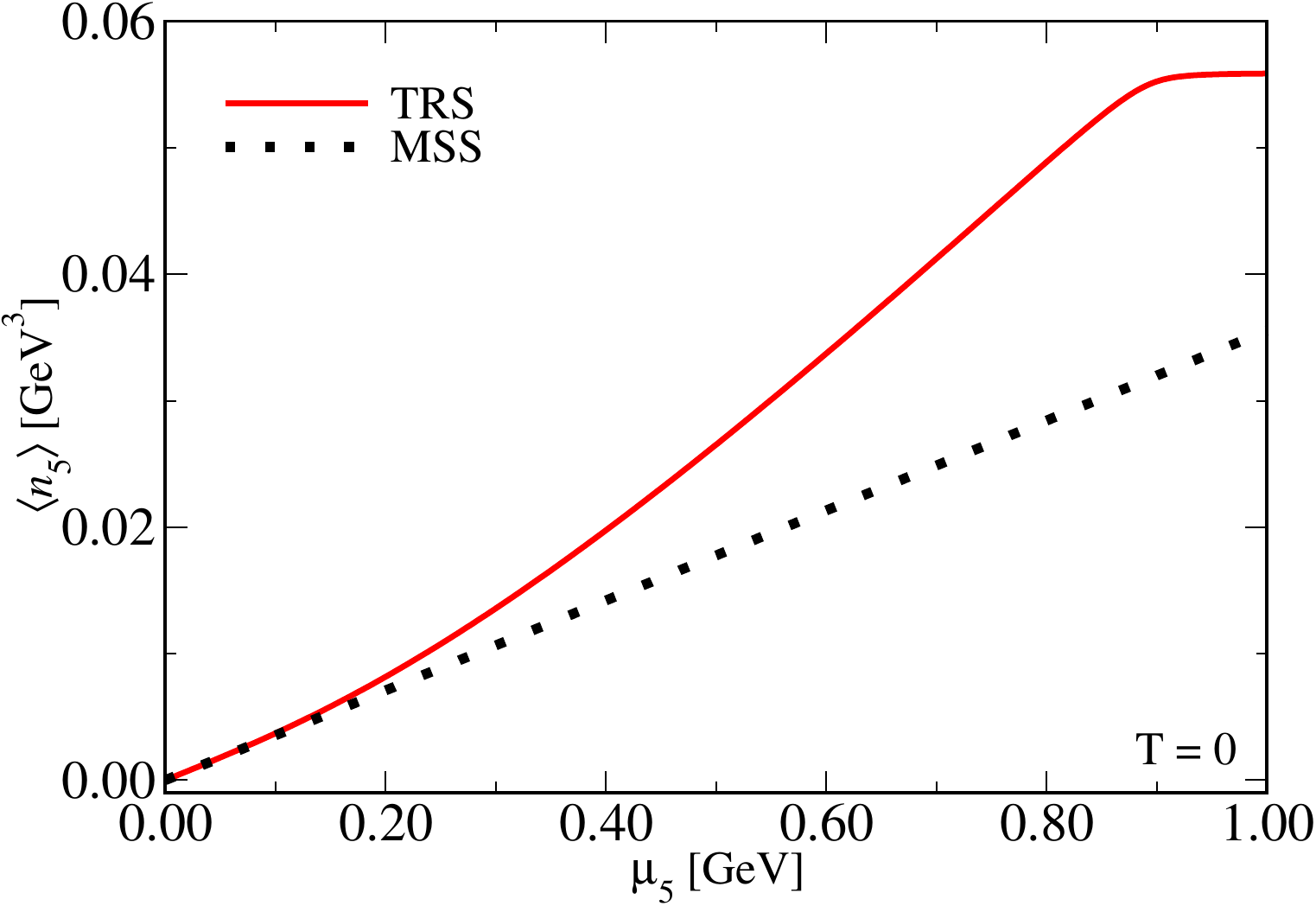} 
 \caption{Chiral density from Eq.~\eqref{n5full} as a function of
   $\mu_5$ at $T = 0$, comparing TRS (left) and MSS (right).}
 \label{Fig3}
 \end{figure}

In {}Fig.~\ref{Fig3} we show the TRS and MSS results for $\langle n_5\rangle\times
\mu_5$ at zero temperature. Although it is clear that in this limit
the contributions of the Polyakov loop vanish, it is interesting
to see that the TRS predicts a plateau for large values of
$\mu_5$. However, with the MSS we obtain a linear increase of the
chiral density with $\mu_5$, which is consistent with ChPT predictions
and lattice QCD simulation results for heavy
pions~\cite{Astrakhantsev:2019wnp}. Although the NJL model
parameters used in this work were determined at the physical value of
$m_{\pi}$, the linear behavior in MSS is obtained directly from the
derivative of Eq.~\eqref{Vmssv2} with respect to $\mu_5$ and it is
not affected by different parametrizations of the model.  The linear
behavior of $\langle n_5\rangle$ as a function of $\mu_5$ and the correct prediction
of the slope when using the MSS are some of the main results of this paper. In the next section we will explore more thoroughly the thermodynamics of the model and the differences produced by TRS and MSS.

\section{Numerical Results at zero quark chemical potential}
\label{sec3}

\begin{center}
\begin{figure}[!htpb]
{\includegraphics[width=8.2cm]{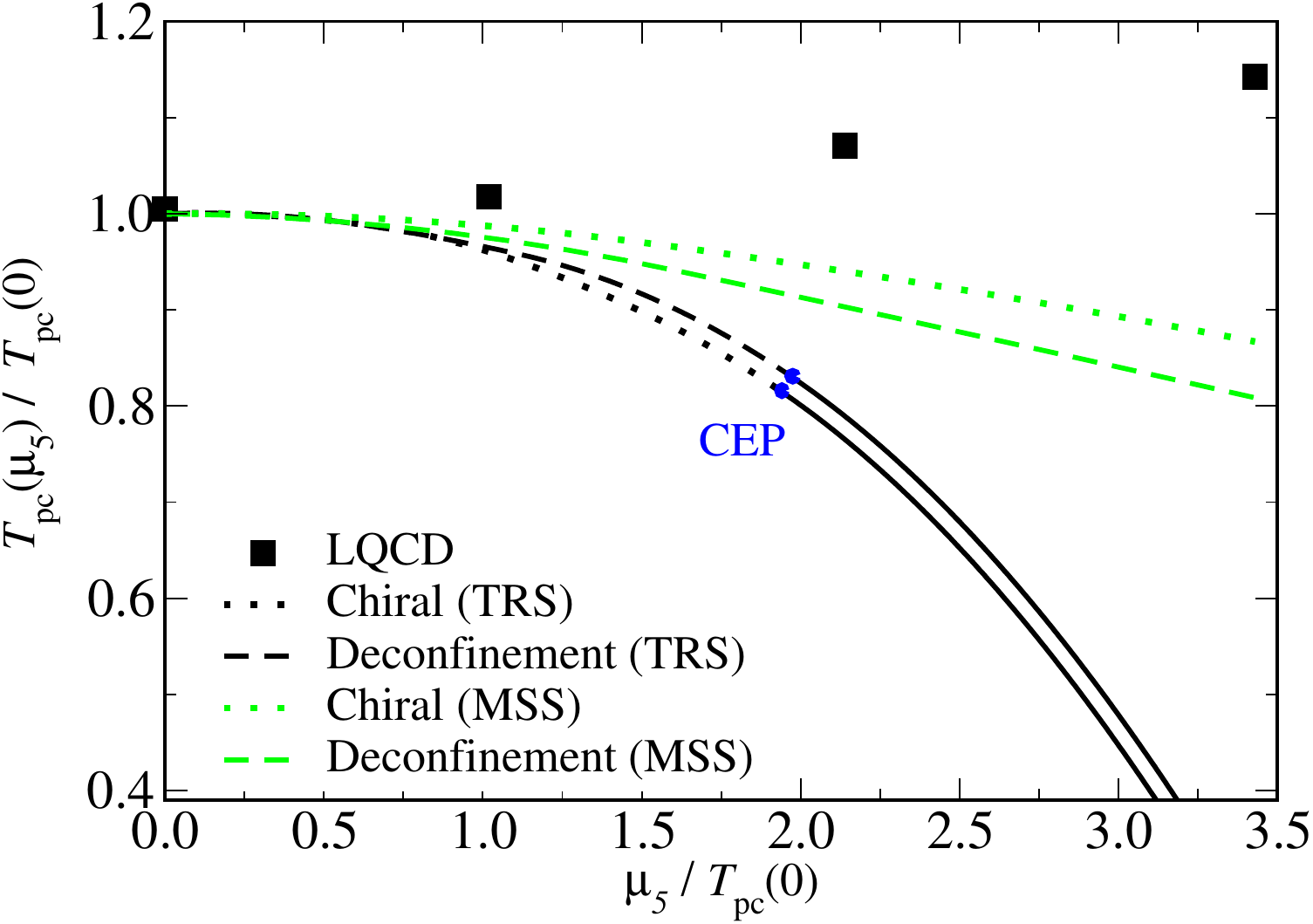}}
 \caption{Phase diagrams in the plane $T\times \mu_5$ for chiral
   restoration and deconfinement, comparing lattice results from
   Ref.~\cite{Braguta:2015owi}, TRS and MSS. The TRS predicts a smooth
   crossover for both transitions for $\mu_5 \lesssim 2T_{\rm pc}(0)$,
   represented by a dotted (chiral) and dashed (deconfinement) lines,
   succeeding by a first-order transition represented by the solid
   lines. The MSS results do not present a CEP, but both
   pseudocritical temperatures are decreasing functions of $\mu_5$.}
 \label{Fig4}
 \end{figure}
\end{center}

In the PNJL model at $\mu = 0$, we have two different (pseudo)critical
temperatures, $T_{\rm pc}^c$ and $T_{\rm pc}^d$, for the chiral and
deconfinement transitions, respectively.  {}For both cases, the
symmetry is only partially restored (the transitions are crossovers), with the (pseudo)critical values determined by the concavity change of the curves, i.e., by the position of the peak of the first derivatives of $M$ and $\Phi$ with respect to $T$: 
\begin{eqnarray}
 \left.\frac{\partial^2 M}{\partial T^2}\right|_{T = T_{\rm pc}^c} =
 \left.\frac{\partial^2\Phi}{\partial T^2}\right|_{T = T_{\rm pc}^d} =
 0.
 \label{Tpc}   
\end{eqnarray}
Recall that the correct order parameter for the chiral transition is the
quark condensate, $\langle\bar q q \rangle$. However, given its linear
relation with the effective quark mass, one may analyze the chiral
symmetry restoration through the evolution of $M$ with the temperature
and/or with the chemical potential without loss of generality.  The
phase diagrams in the $T\times \mu_5$ plane for the chiral and deconfinement transitions are shown in
{}Fig.~\ref{Fig4}.  The values of the pseudocritical temperatures at $\mu_5 = 0$, $T_{\rm pc}(0)$, are given in
Table~\ref{Tpc0} for both regularization schemes.

\begin{table}[ht!]
\caption{\label{Tpc0} Values of pseudocritical temperatures for chiral
  ($T_{\rm pc}^c$) and deconfinement ($T_{\rm pc}^d$) phase
  transitions.}
\begin{center}
\begin{tabular}{ccc}
\hline \hline & \hspace*{.4cm}$T_{\rm pc}^c$ (GeV)\hspace*{.4cm}
& \hspace*{.4cm}$T_{\rm pc}^d$ (GeV) \hspace*{.4cm} \\ \hline  TRS &
0.229 & 0.225 \\ MSS & 0.233 & 0.227 \\\hline \hline
\end{tabular}
\end{center}
\end{table}

Recent lattice simulations predict a continuous and smooth {\it increasing} behavior of the order parameters with $\mu_5$ for both the chiral and deconfinement transitions. In addition, the values for $T_{\rm
  pc}^c$ and $T_{\rm pc}^d$ are approximately the
same~\cite{Braguta:2015owi}. These results are represented by the
squared dots in {}Fig.~\ref{Fig4}.  However, one can see that the PNJL model in the presence of a chiral imbalance using the standard
form for the fit of Polyakov loop, given by Eq.~\eqref{Uphi},
is not able to reproduce the lattice QCD behavior. Although the MSS does
not predict a critical end point (CEP) as the standard NJL model does~\cite{Farias:2016let}, both schemes lead to
decreasing functions of the pseudocritical temperatures with $\mu_5$. We have also observed a similar behavior with other  Polyakov-loop parametrizations commonly considered in the literature. In all  available parametrizations, even when using the MSS the pseudocritical
temperatures are always decreasing functions of the chiral chemical
potential. Our proposal here is that, in order to obtain results in agreement with the lattice results for the pseudocritical temperatures,
these parametrizations need to be changed so that they not only
depend on the temperature, but also depend on $\mu_5$.  Thus, we
propose an explicit redefinition of the Polyakov-loop potential such
as to include an explicit dependence of this function with $\mu_5$,
i.e.,  $\mathcal{U}(\Phi,T) \to \mathcal{U}(\Phi,T,\mu_5)$.\footnote{As
we have already mentioned previously, $\Phi = \Phi^{\dagger}$ at $\mu
= 0$.} Still working with the parametrization of the form in
Eq.~\eqref{V1pnjl}, we propose modifying the coefficient of  $\Phi\Phi^{\dagger}$ in the potential $\mathcal{U}$ as follows:
\begin{eqnarray}
 \mathcal{U}(\Phi,T,\mu_{5})&=&T^{4}\bigg[-\frac{\bar{b}_{2}(T,\mu_{5})}{2}\Phi\Phi^{\dagger}
   \nonumber\\ &-& \frac{b_{3}}{6}(\Phi^{3} +
               {\Phi^{\dagger}}^{3})+\frac{b_{4}}{4}(\Phi
               \Phi^{\dagger})^{2}\bigg],\label{Uphi2}  
\end{eqnarray}
in which
\begin{eqnarray}
 \bar{b}_{2}(T,\mu_{5})&=&a_{0}+a_{1}\left(\frac{T_{0}}{T}\right)+a_{2}\left(\frac{T_{0}}{T}\right)^{2}+a_{3}\left(\frac{T_{0}}{T}\right)^{3}
 \nonumber\\ &+&k_1\left(\frac{\mu_{5}}{T}\right) +
 k_2\left(\frac{\mu_{5}}{T}\right)^{2}+k_3\left(\frac{\mu_{5}}{T}\right)^{3}\,.\label{b2mu5}   
\end{eqnarray}
To obtain the MSS results presented in the remainder of this paper we
have used $k_1 = 0.08602,k_2 = -1.39646$ and $k_3 = 0.3381$.  These
values of constants can be seen to be the fitting values needed to reproduce the lattice results when a chiral chemical potential is present.

Traditional implementations of the Polyakov loop in quark models typically rely on fitting lattice results for a gluon gas at finite temperature. However, these models often fail to accurately replicate the qualitative behavior observed in lattice QCD simulations, indicating that both the quark and gauge sectors are influenced by medium effects beyond temperature. To address this discrepancy, it is common to introduce additional dependencies into the Polyakov-loop potential to align with lattice data. 
{}For example, the authors in Ref.~\cite{Schaefer:2007pw} proposed a dependence on quark chemical potential $\mu$ and the number of flavors, employing renormalization group arguments to investigate the thermodynamic properties of the quark-meson model. Similarly, Ref.~\cite{Adhikari:2018cea} modified the dependence on $\mu$ by introducing the isospin chemical potential $\mu_I$. In Ref.~\cite{Ferreira:2013tba} it was incorporated an implicit dependence on the magnetic field in the entangled PNJL (EPNJL) model, where the coupling constant is modified to depend on the order parameters $\Phi$ and $\bar \Phi$. The resulting model then predicted inverse magnetic catalysis, a feature observed in lattice simulations but not in the ordinary PNJL approach. The authors of Ref.~\cite{Sun:2024anu} studied rotating systems by including a dependence on angular velocity $\omega$ in $\cal U$ and $\Phi$, finding that critical temperatures increase with $\omega$ for both chiral and deconfinement transitions, consistent with lattice data. Then, in Ref.~\cite{Dexheimer:2009hi} it was proposed a dependence of the chemical potential on the Polyakov-loop potential that persists at $T = 0$, enabling the study of density-driven deconfinement. This possibility is particularly relevant in light of recent gravitational wave data analyses.

In this study, we adopted a polynomial dependence of the coefficient $\bar b_2$ on $\mu_5$, analogous to its existing dependence on $T$ as shown in Eq.~(\ref{b2mu5}). We acknowledge that numerous alternative parametrizations could be employed instead of the chosen polynomial. However, we opted for the simplest approach to demonstrate that the key missing element for these models to accurately replicate lattice QCD simulations is a dependence of the Polyakov-loop potential on $\mu_5$, in addition to its dependence on $T$.

\begin{center}
 \begin{figure}[htpb!]
\subfigure[]{\includegraphics[scale=0.33]{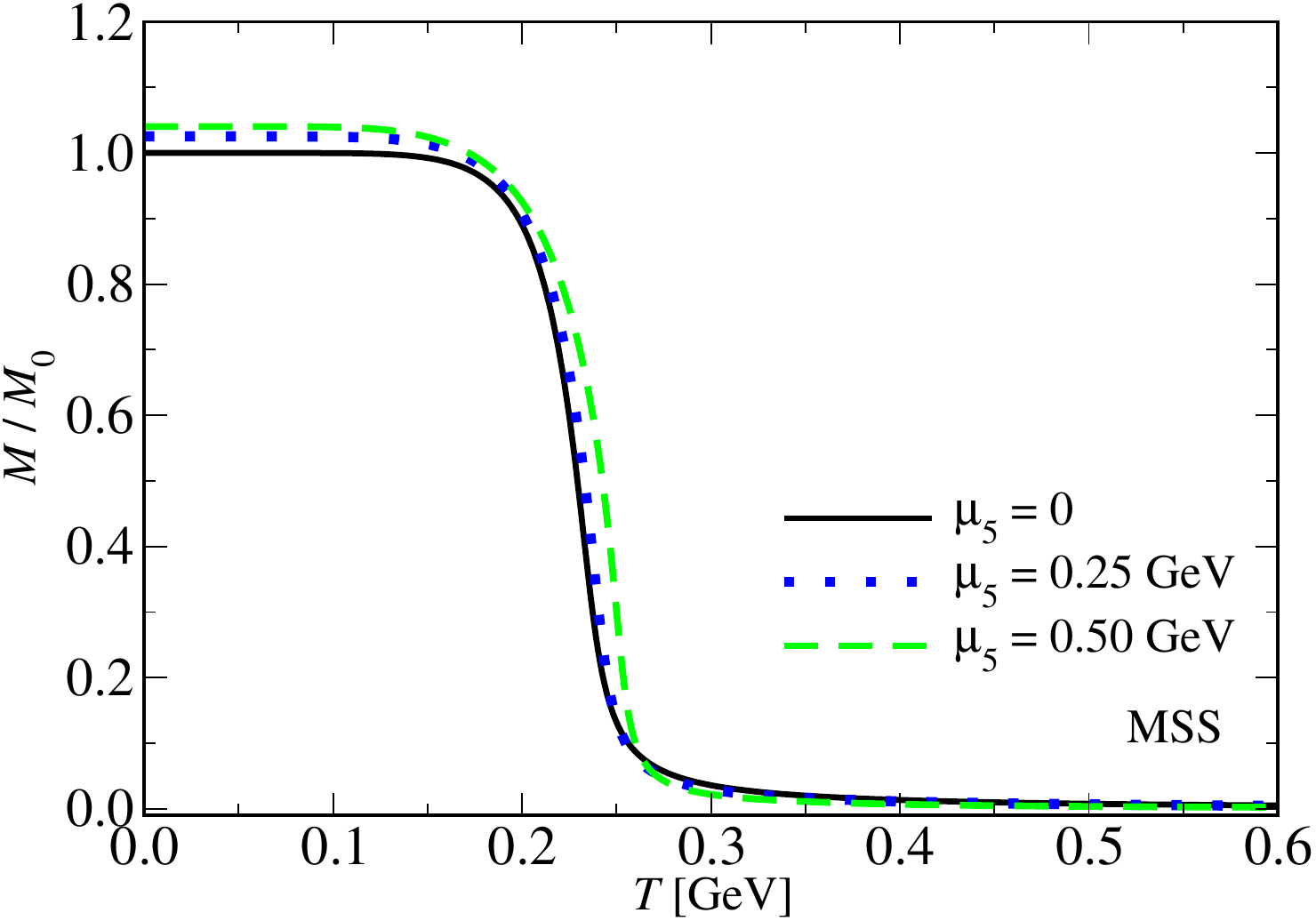}}
\subfigure[]{\includegraphics[scale=0.33]{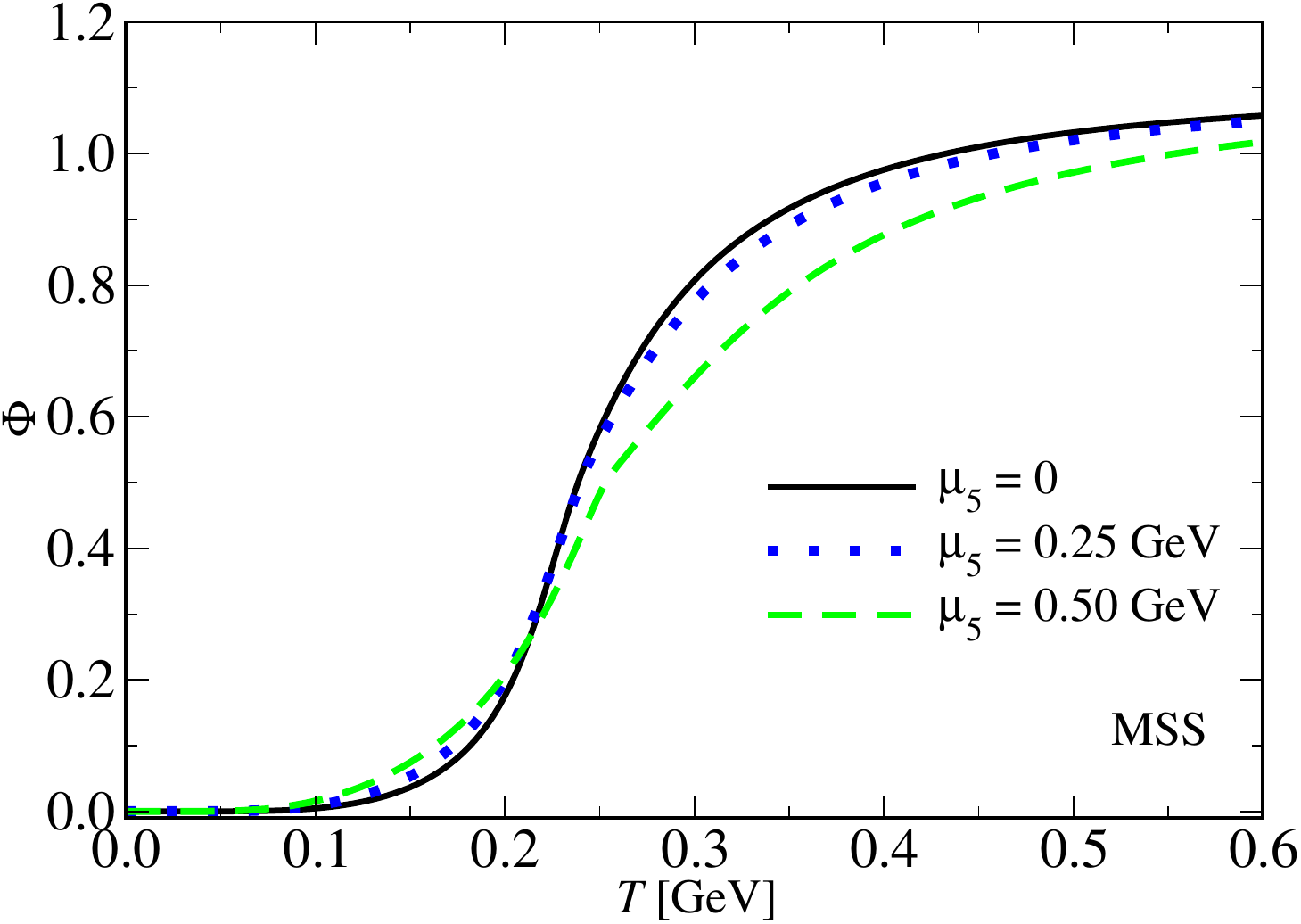}}
\caption{MSS results for the normalized quark mass $M/M_{0}$ on panel (a) and Polyakov loop $\Phi$ on panel (b) as functions of the temperature, for $\mu = 0$ and different values of $\mu_5$. Results obtained with the redefined Polyakov-loop potential as given by Eq.~\eqref{Uphi2}. }
\label{Fig5}
\end{figure}
\end{center} 

\begin{figure}[htpb!]
\begin{center}
\includegraphics[scale=0.33]{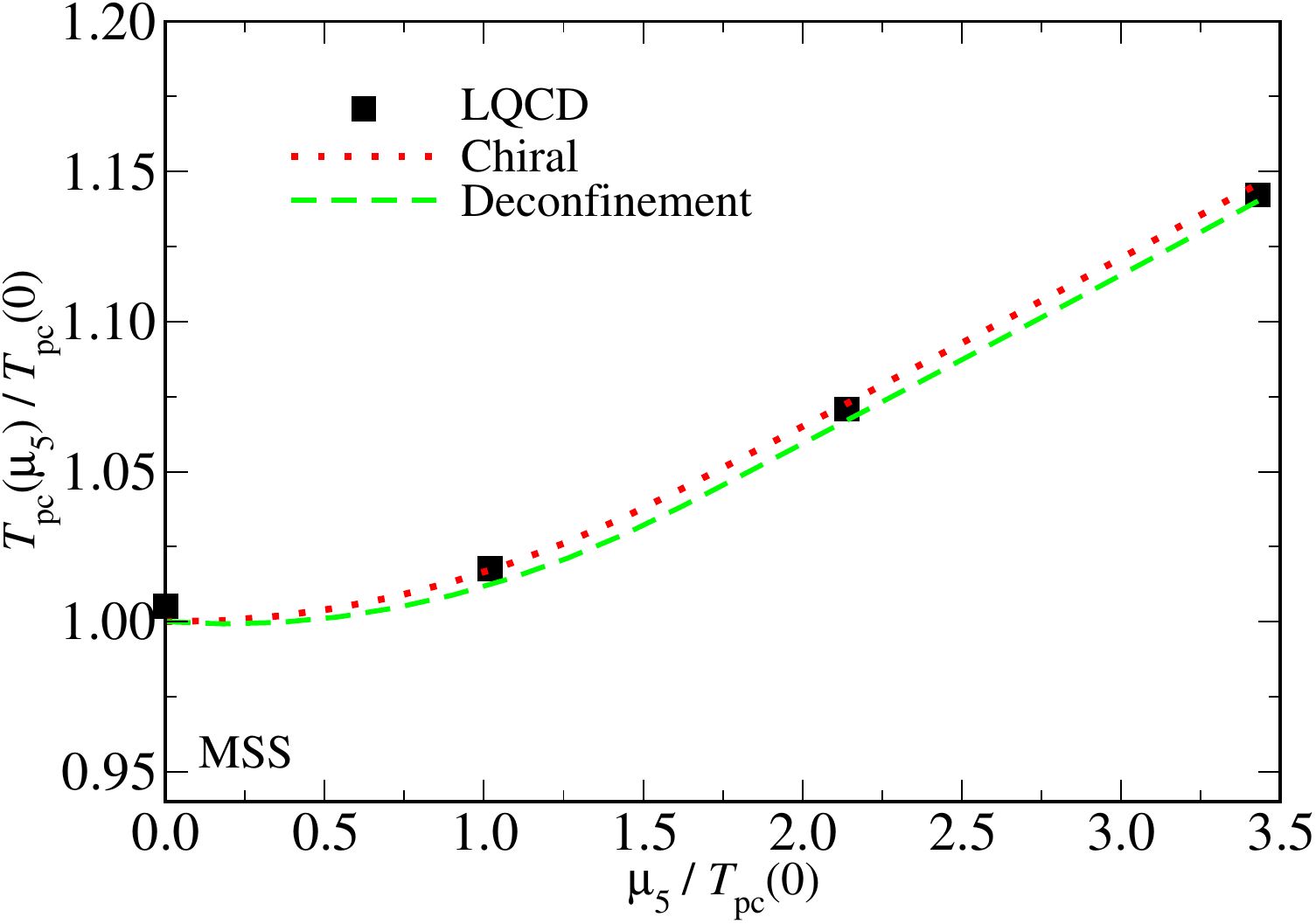}
\caption{Phase diagrams in the plane $T\times \mu_5$ for chiral
  restoration and deconfinement, comparing lattice results from
  Ref.~\cite{Braguta:2015owi} and MSS, including a dependence with
  $\mu_5$ on the Polyakov-loop potential, as given in
  Eq.~\eqref{Uphi2}. The dotted and dashed lines are the MSS results for
  chiral and deconfinement transitions, respectively, while large dots
  are lattice results.}
\label{Fig6}
\end{center}
\end{figure}

In {}Fig.~\ref{Fig5} we show the results for the normalized quark mass
and Polyakov loop $\Phi$ as functions of the temperature for the MSS,
using the newly redefined $\mathcal{U}(\Phi,T,\mu_{5})$ as given in
Eq.~\eqref{Uphi2}. {}From this figure one can see that the inflection
points in the curves are shifted to the right, which means that the
pseudocritical temperatures now increase with $\mu_5$. This is
the opposite behavior to that observed in {}Fig.~\ref{Fig2}. The
results for the pseudocritical temperatures as a function of  $\mu_5$,
when using Eq.~\eqref{Uphi2}, are shown in {}Fig.~\ref{Fig6}. {}From
{}Fig.~\ref{Fig6}  one can see that the dependence of the Polyakov
potential with the chiral chemical potential enables the PNJL model to
reproduce the correct behavior of lattice simulation results and also
make the values of $T_{\rm pc}^c$ and $T_{\rm pc}^d$ closer to each
other.  Note that if we consider the TRS method instead and by using
$\mathcal{U}(\Phi, T,\mu_5)$ , we still obtain a CEP in the $T\times
\mu_5$ plane, which is not consistent with the lattice simulations;
for this reason, we opted for showing only the MSS results in this
case.

\subsection{Thermodynamic quantities}

Next, we study the thermodynamics of the PNJL model at
$\mu = 0$, comparing the TRS with $\mathcal{U}(\Phi,T)$ and MSS with
$\mathcal{U}(\Phi,T,\mu_5)$.  {}From the thermodynamic potential given
in Eq.\eqref{V1pnjl} it is possible to compute  several thermodynamic quantities, starting from the normalized pressure: 
\begin{align}
 p_N(M,\bar{M},\Phi,\Phi^{\dagger},T,\mu,\mu_5) =  &-
 \bigg[\Omega(M,\Phi,\Phi^{\dagger},T,\mu,\mu_5)
   \nonumber\\ &-
   \Omega(\bar{M},0,\mu,\mu_5)\bigg],
 \label{pressN}
\end{align}
where $\bar{M}\equiv M(\mu,\mu_5)$ is defined at finite chemical
potentials but zero temperature\footnote{From Eq.~\eqref{Uphi} one can
see that all the $\Phi$ and $\Phi^{\dagger}$ dependence with the
temperature vanishes in the thermodynamic potential at $T = 0$.}. {}For
simplicity, we will omit the functional dependencies in all thermodynamic quantities. The entropy $s$ and energy $\varepsilon$ densities are defined at fixed $\mu_5$, respectively as 
\begin{align}
& s =  \left.\frac{\partial p_N}{\partial
    T}\right|_{\mu_5},\label{entropy}\\ & \varepsilon = T s - p_N +
  \mu_5 \langle n_{5_N}\rangle,
\label{epsilon}
\end{align}
where the normalized chiral density $\langle n_{5_N}\rangle$ is defined in terms of
the normalized pressure as 
\begin{equation}
\langle n_{5_N}\rangle = \frac{\partial p_N}{\partial\mu_5}.
 \label{n5N}
\end{equation}
Another interesting quantity is the squared speed of sound, also at
finite $\mu_5$,
\begin{equation}
c_{s}^{2} = \left.\frac{\partial p_N}{\partial
  \varepsilon}\right|_{\mu_5} = \frac{s}{C_{v}},
\label{cs2}
\end{equation}  
that may be defined in terms of the specific heat as
\begin{equation}
    C_{v} = T \frac{\partial  s}{\partial T}\,.
    \label{Cv}
\end{equation}
It is common to find in the literature an alternative definition of the squared speed of sound, calculated at fixed entropy per baryon $s/n_B$. This alternative definition is motivated in the context of heavy ion collisions since the speed of sound is approximately constant in the whole expansion stage of the collision~\cite{Braun-Munzinger:2015hba}. Since we are working at zero quark density, we will use the definition of $c_s^2$ at finite chiral chemical potential as given in Eq.~\eqref{cs2}. Finally, we also consider the scaled trace anomaly $\Delta$, also referred to in the literature as `interaction measure':
\begin{equation} 
\Delta = \frac{1}{T^4} \left( \varepsilon - 3p_N\right),
\label{Delta}
\end{equation}
This quantity helps to assess the high temperature limit of the system as provides a measure of the deviation from the scale invariance.

\begin{figure*}[htpb!]
 \subfigure[]{\includegraphics[scale=0.33]{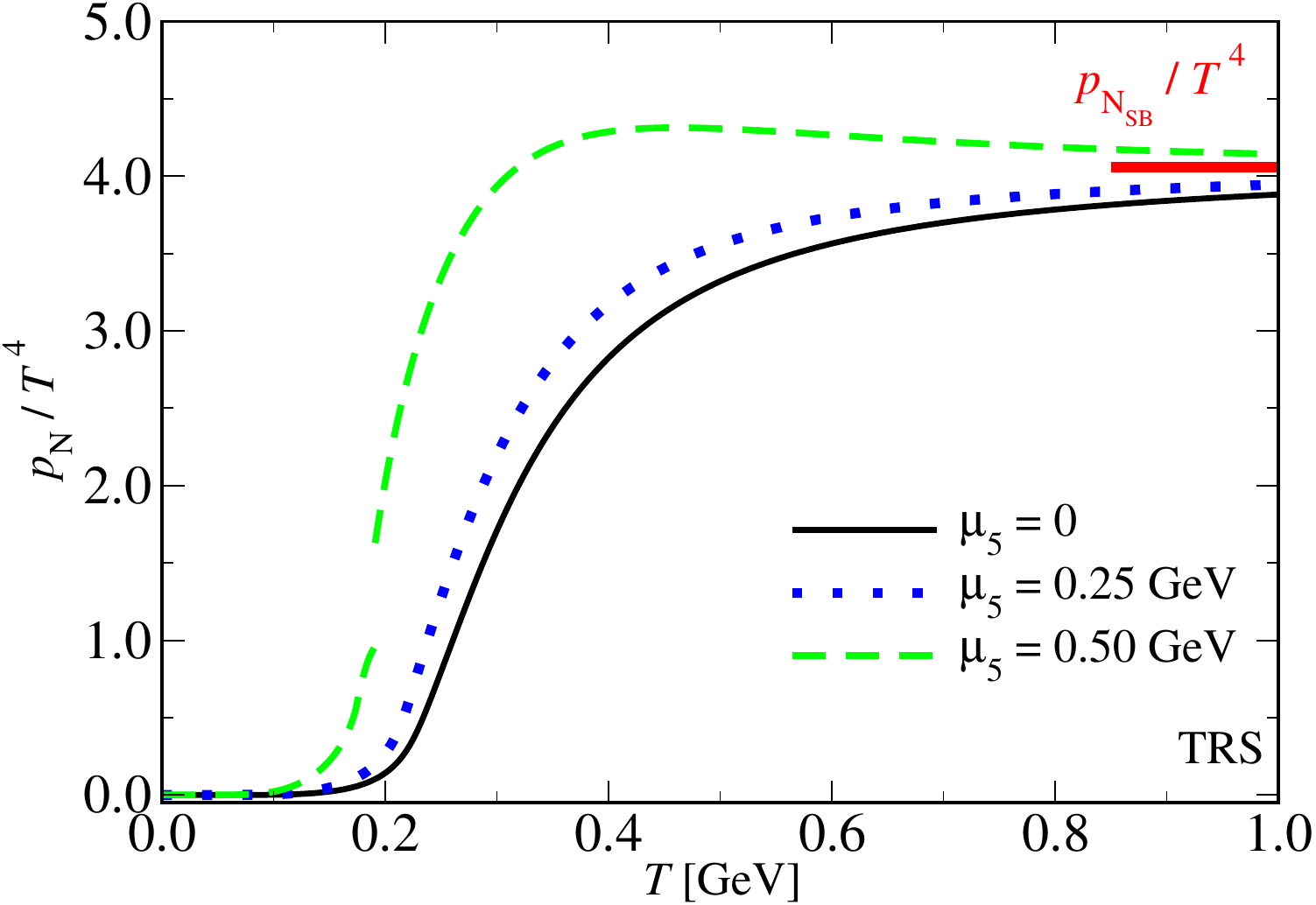}}
 \subfigure[]{\includegraphics[scale=0.33]{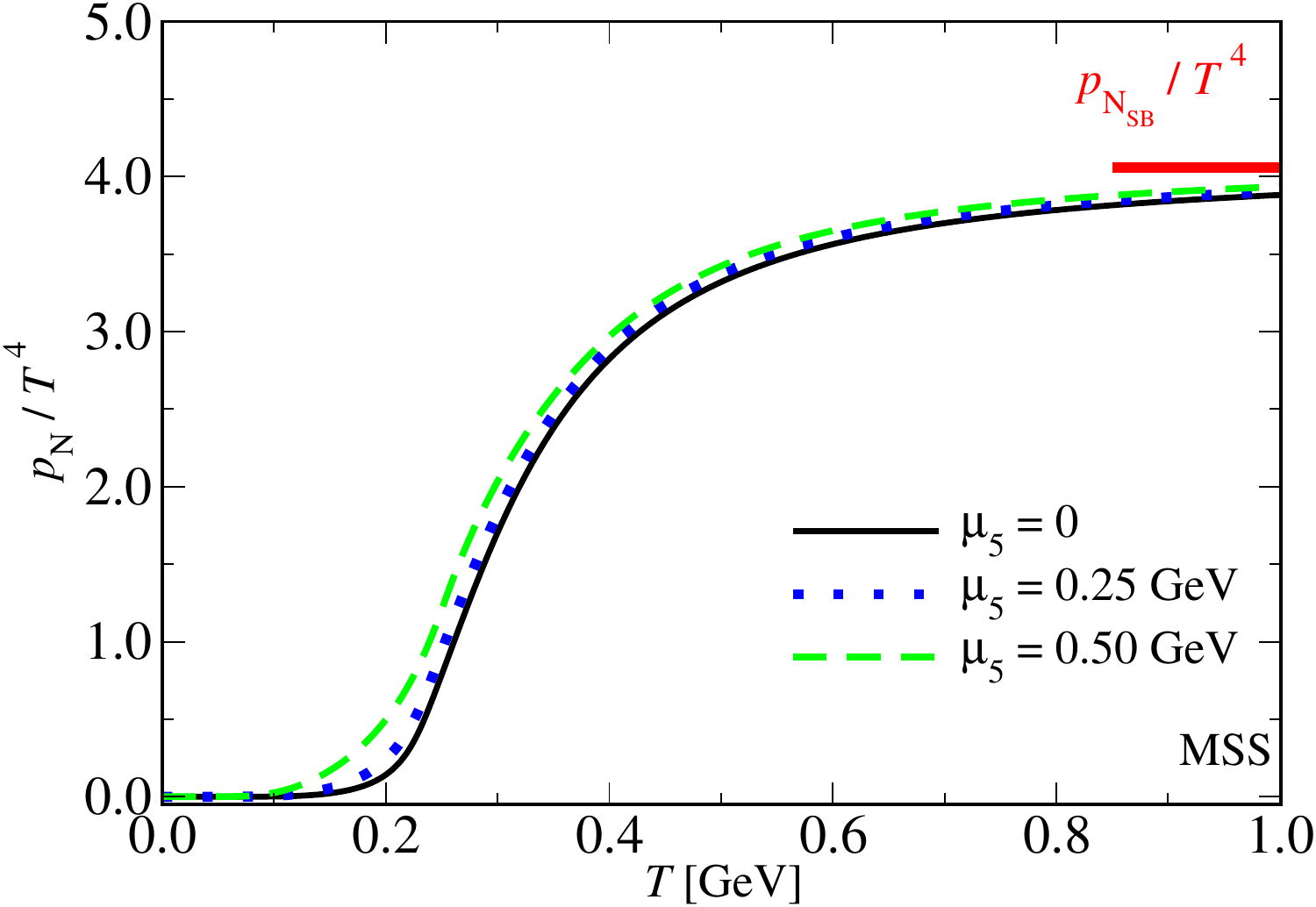}}
 \subfigure[]{\includegraphics[scale=0.33]{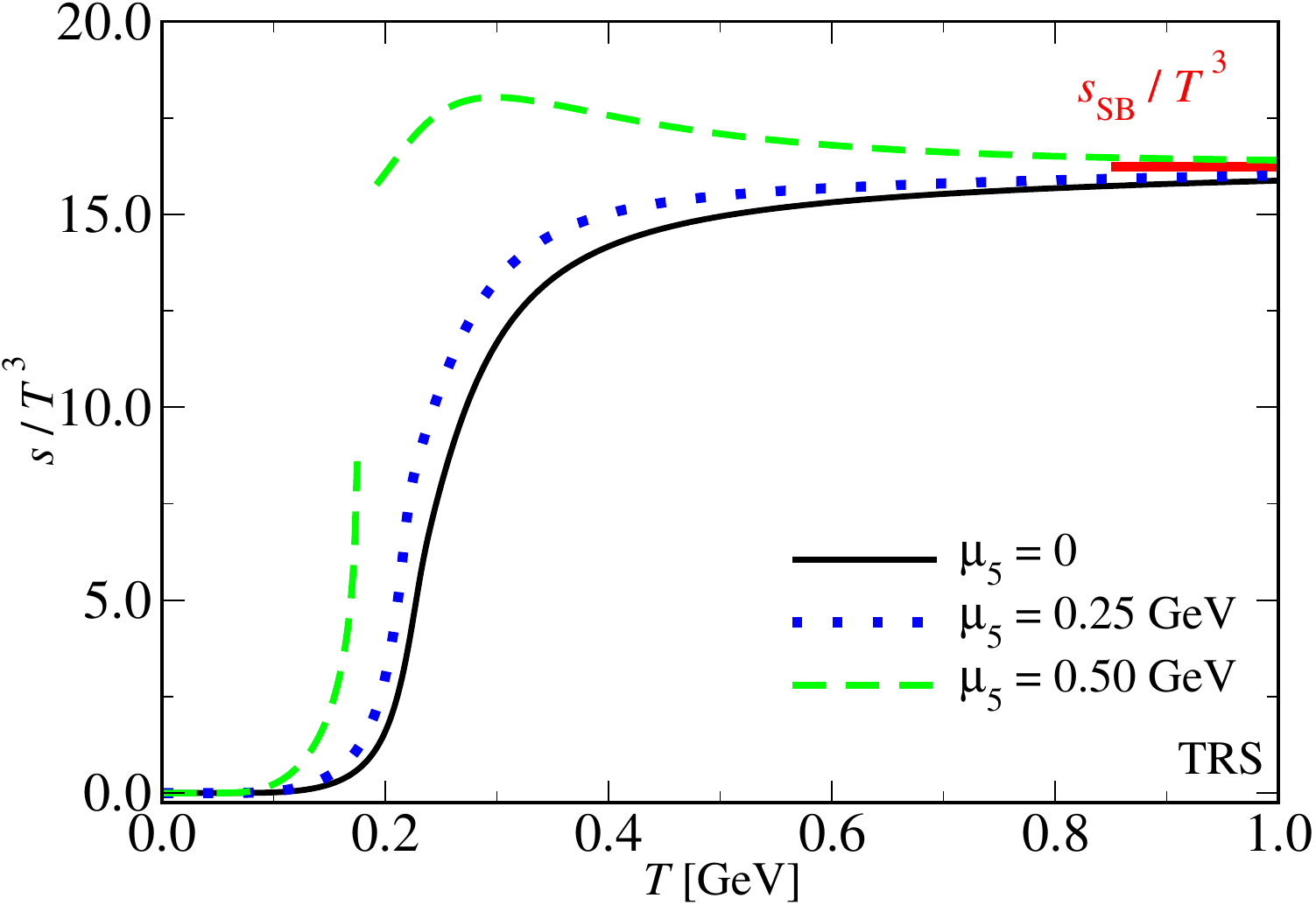}} \subfigure[] {
   \includegraphics[scale=0.33]{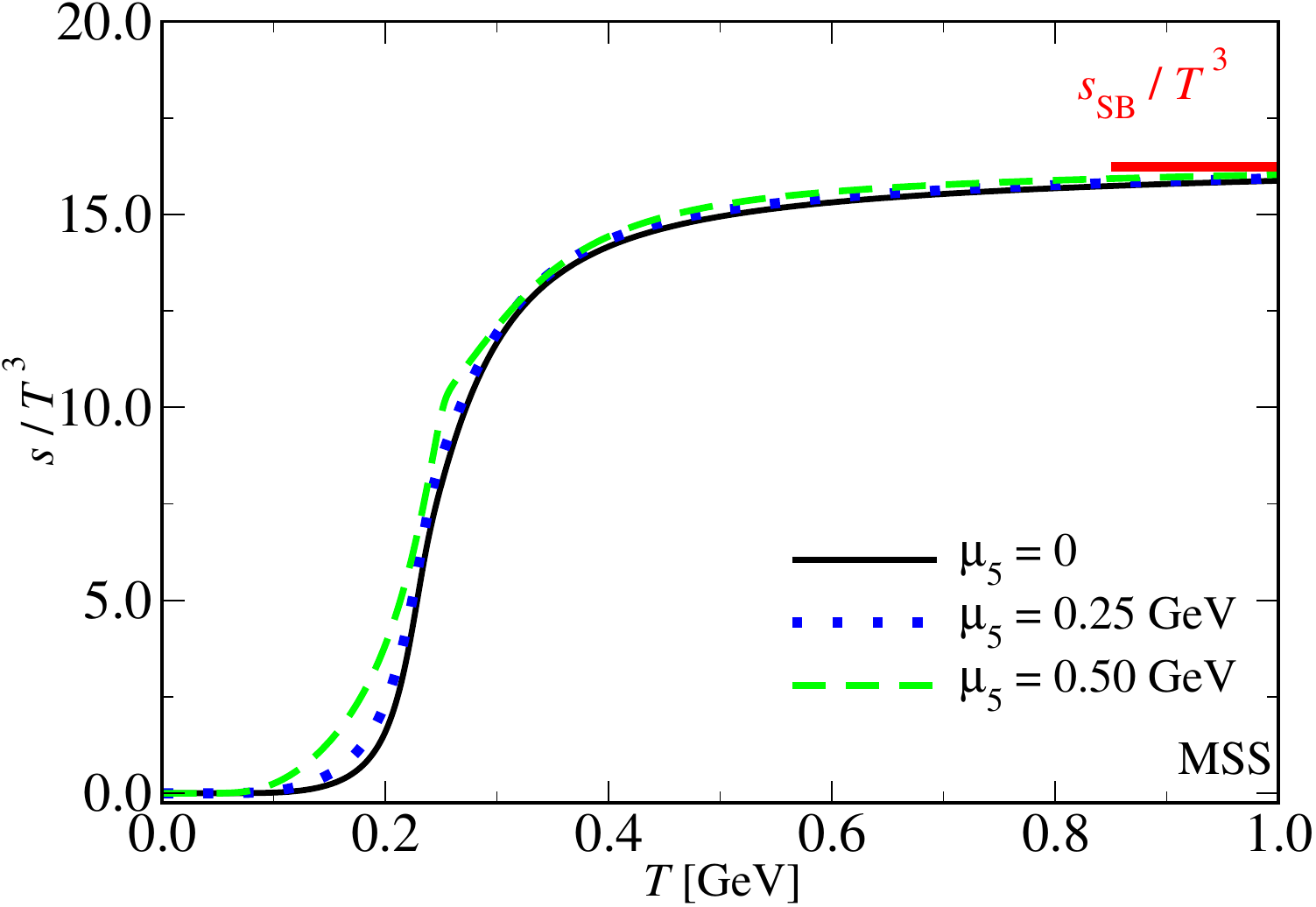}}
 \subfigure[]{\includegraphics[scale=0.33]{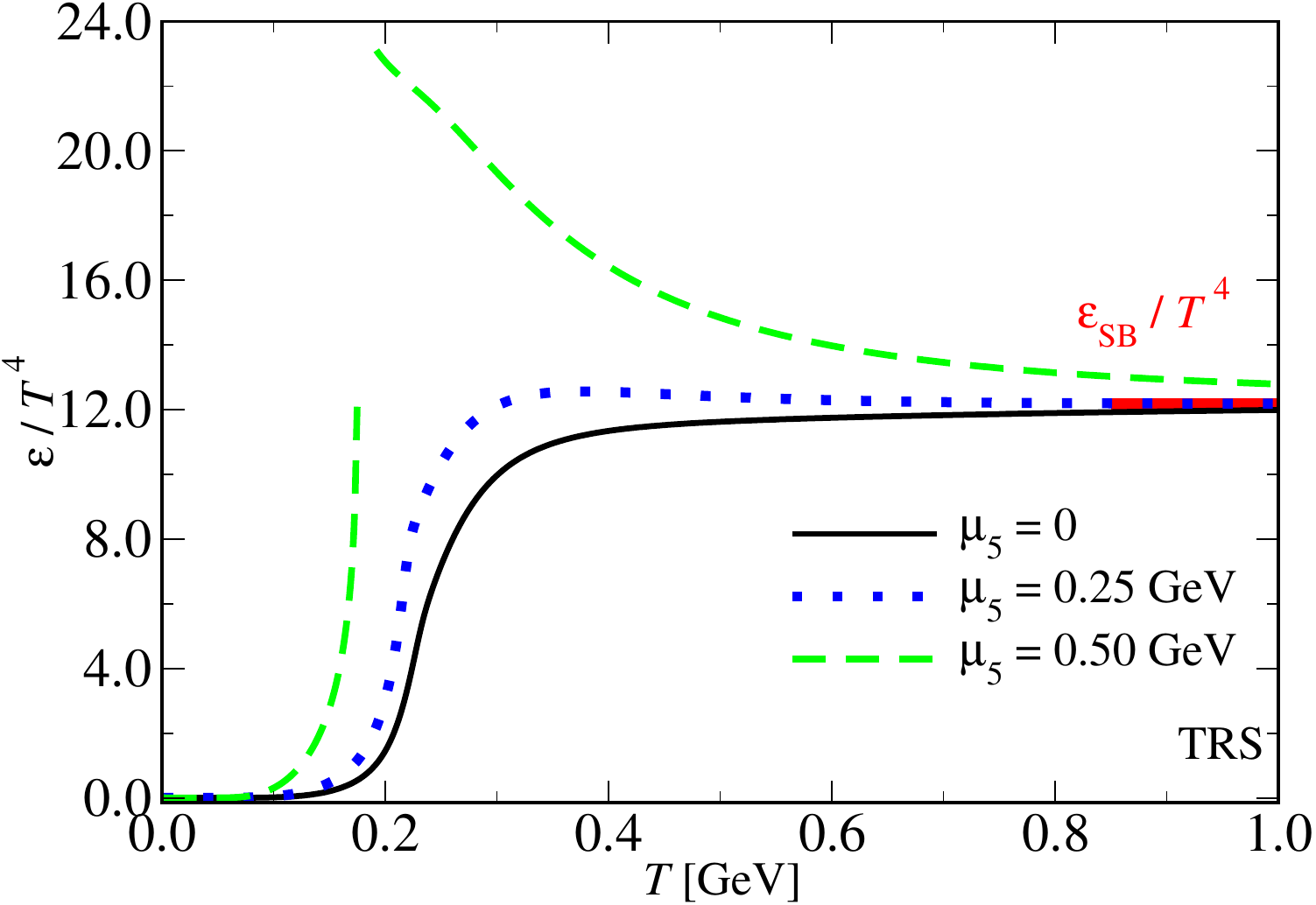}} \subfigure[]{
   \includegraphics[scale=0.33]{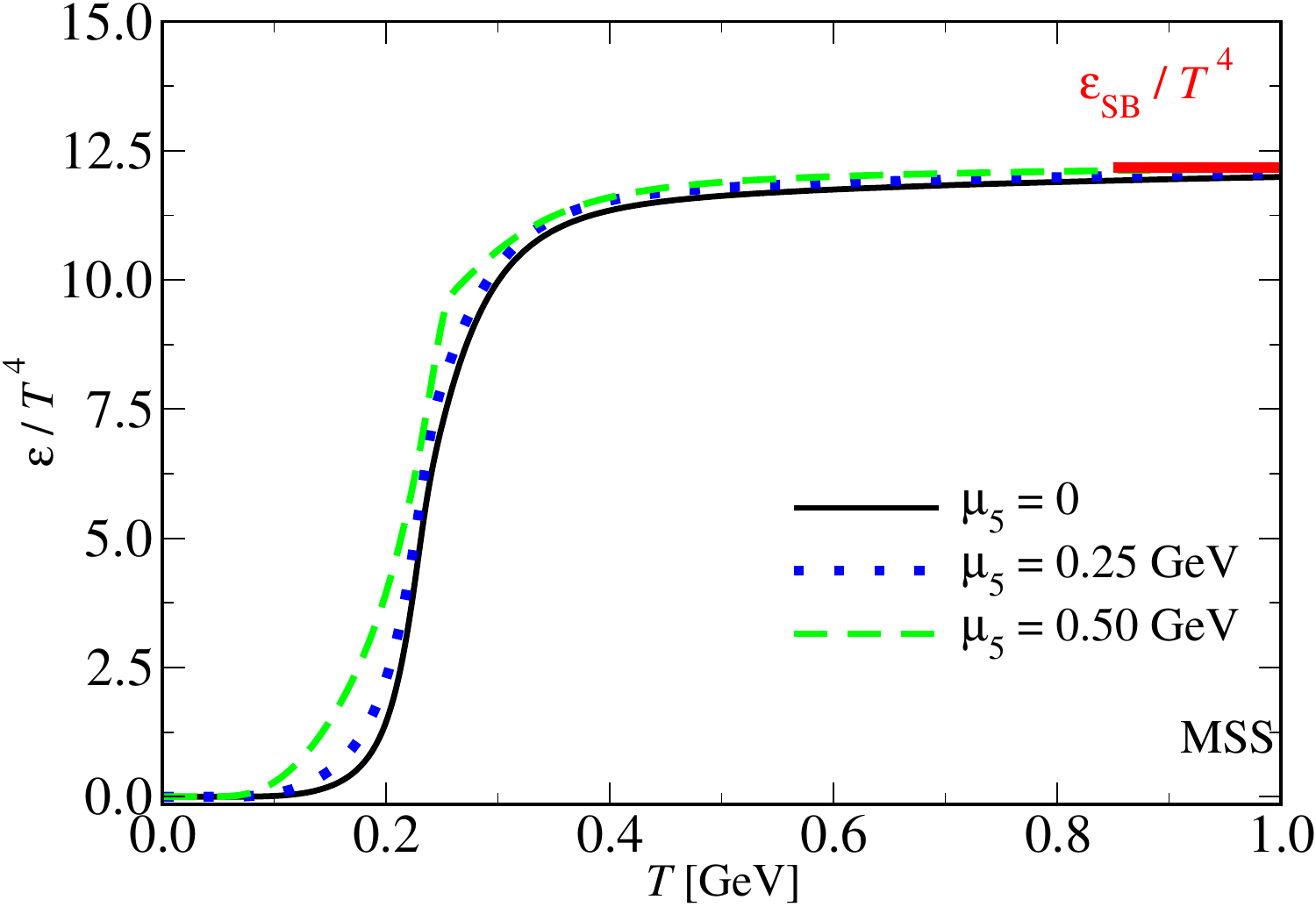}}
\caption{The normalized pressure $p_N/T^4$, entropy $s/T^3$ and energy density 
$\varepsilon/T^4$ as a function of temperature and different values of $\mu_5$ (all 
at $\mu=0$) in the TRS case, panels (a), (c) and (e), respectively and in the MSS 
case,  panels (b), (d) and (f), respectively. All results obtained using 
$\mathcal{U}(\Phi,T,\mu_5)$ as given by Eq.~\eqref{Uphi2}.}
\label{Fig7}
\end{figure*}

\begin{figure*}[htpb!]
\subfigure[]{\includegraphics[scale=0.33]{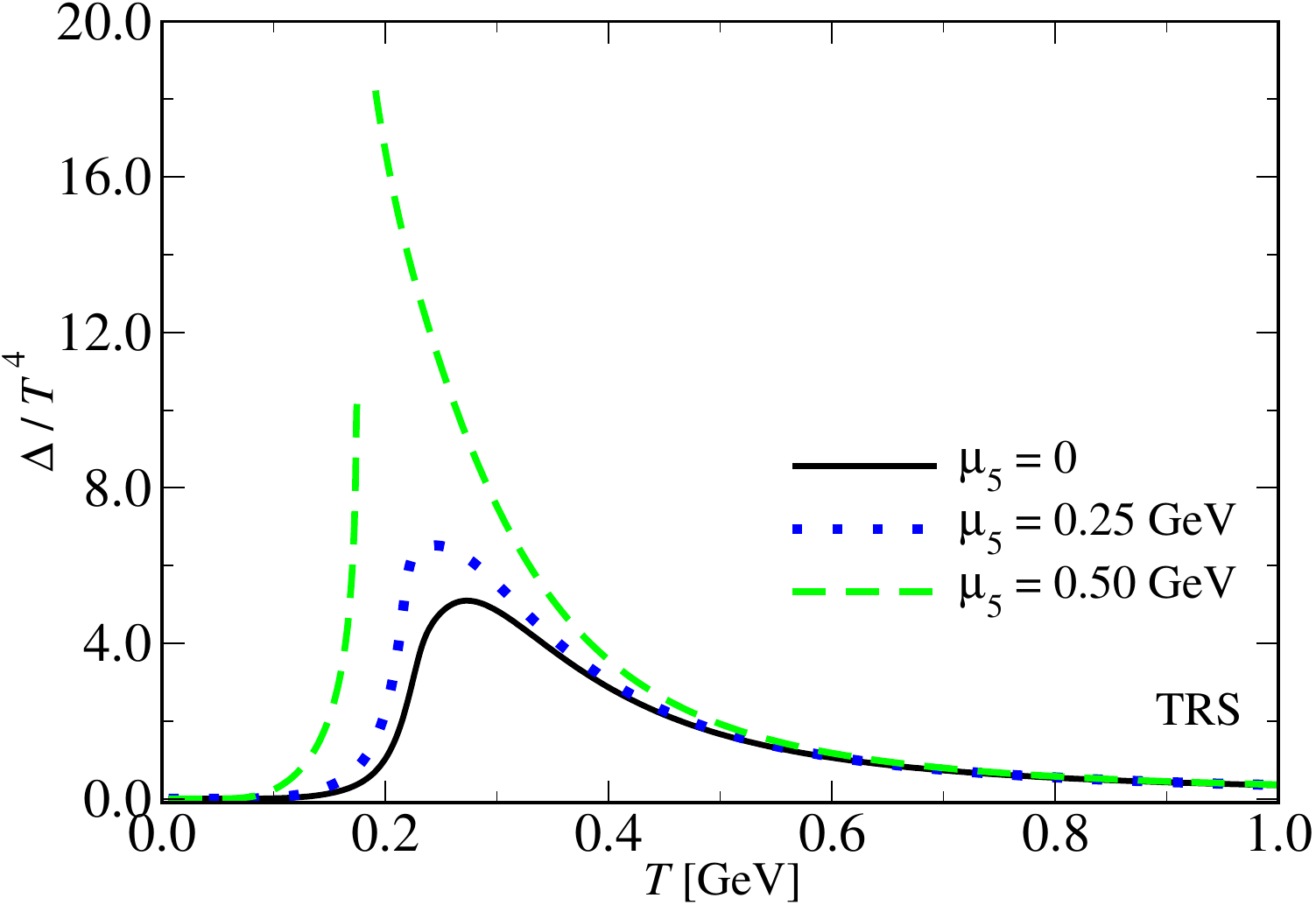}}
\subfigure[]{\includegraphics[scale=0.33]{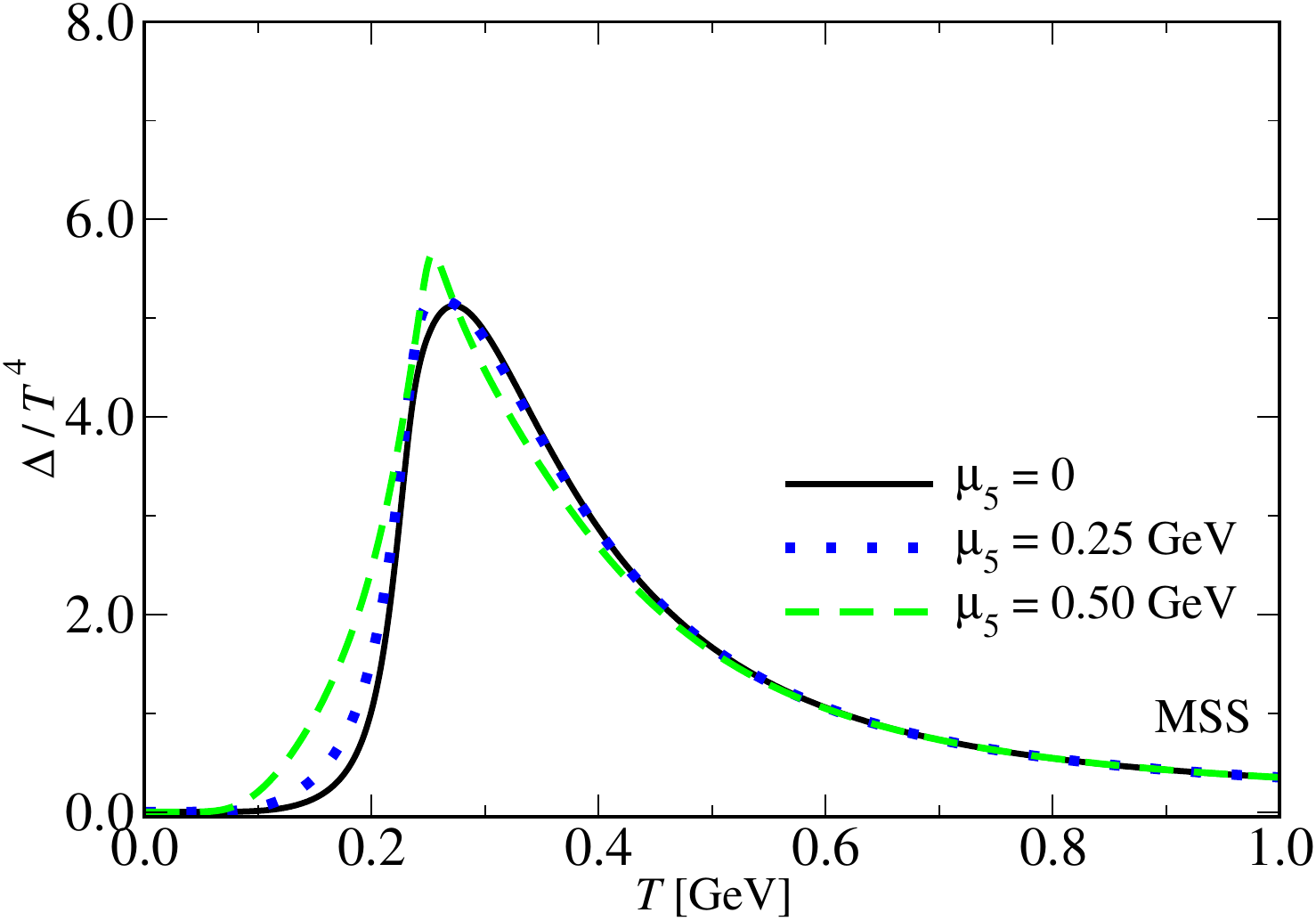}}
\subfigure[]{\includegraphics[scale=0.33]{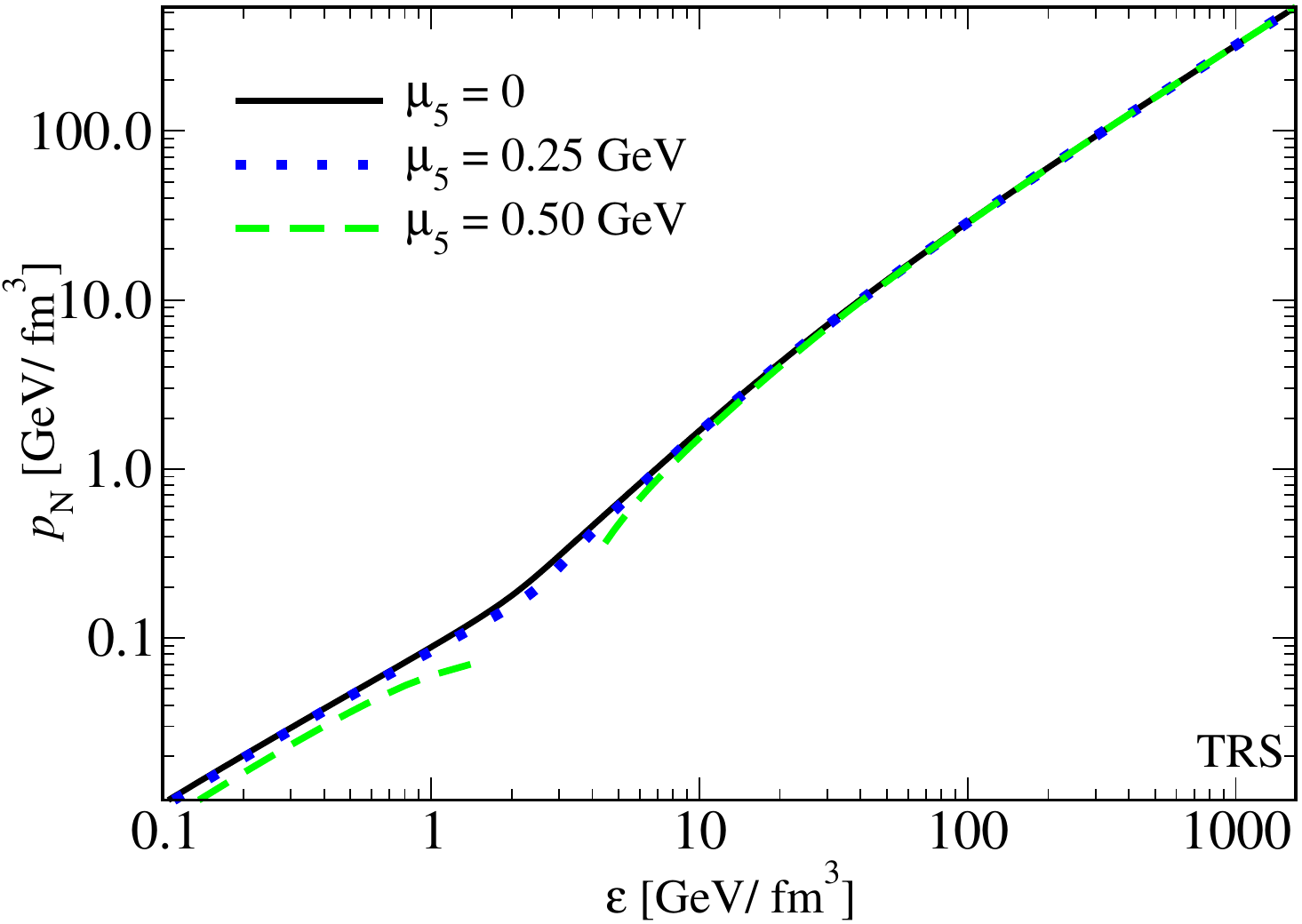}}
\subfigure[]{\includegraphics[scale=0.33]{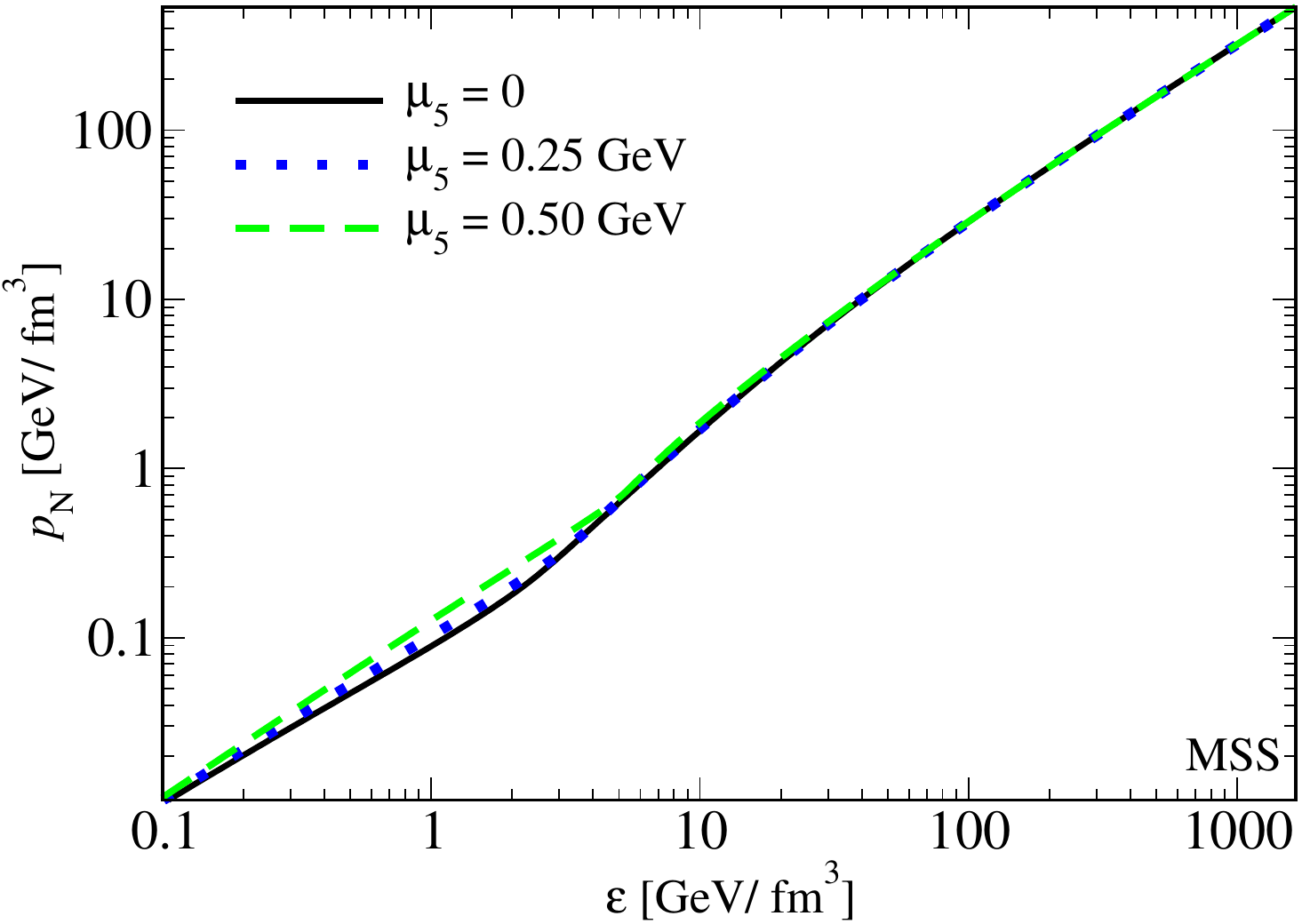}}
\subfigure[]{\includegraphics[scale=0.33]{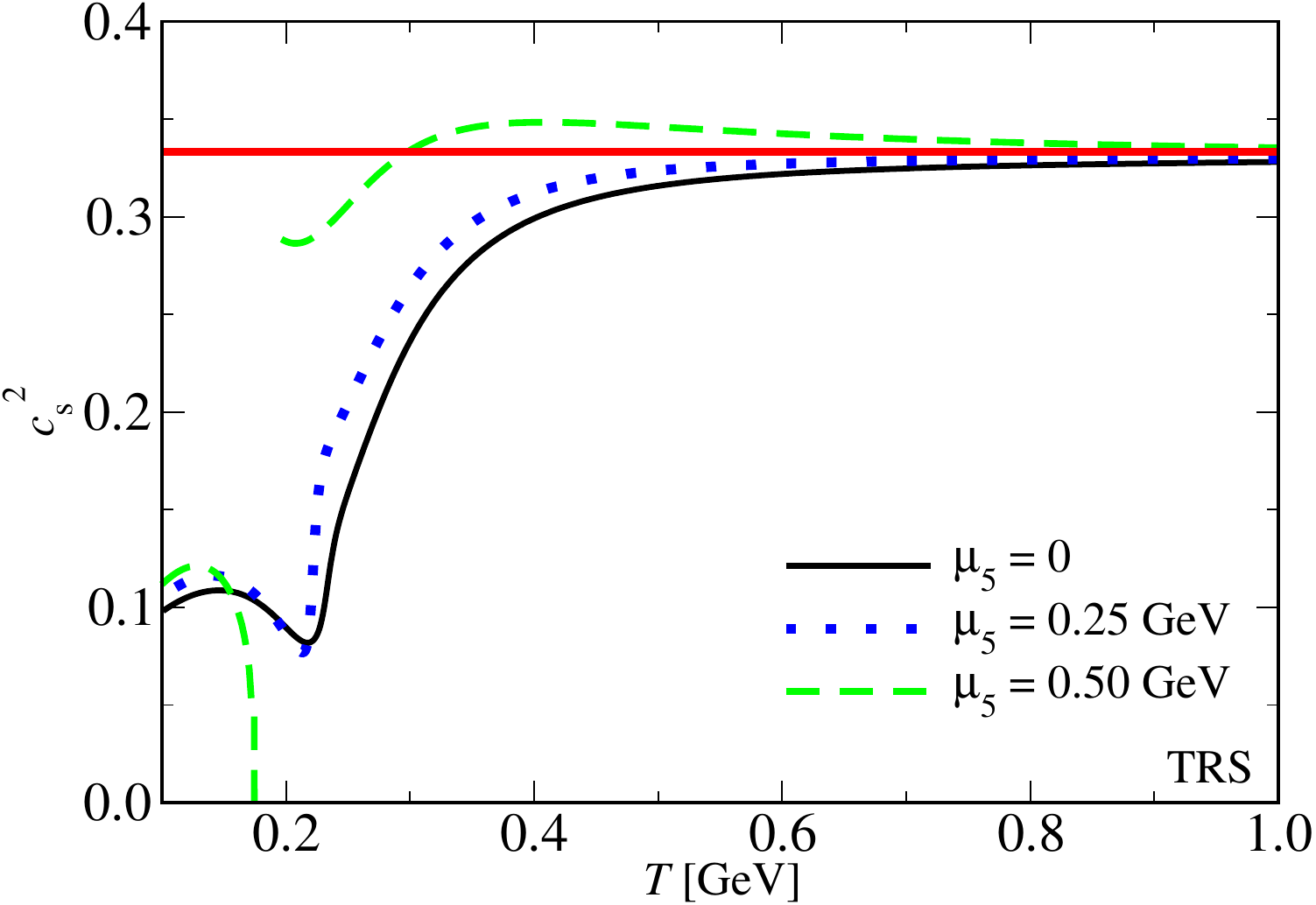}}
\subfigure[]{\includegraphics[scale=0.33]{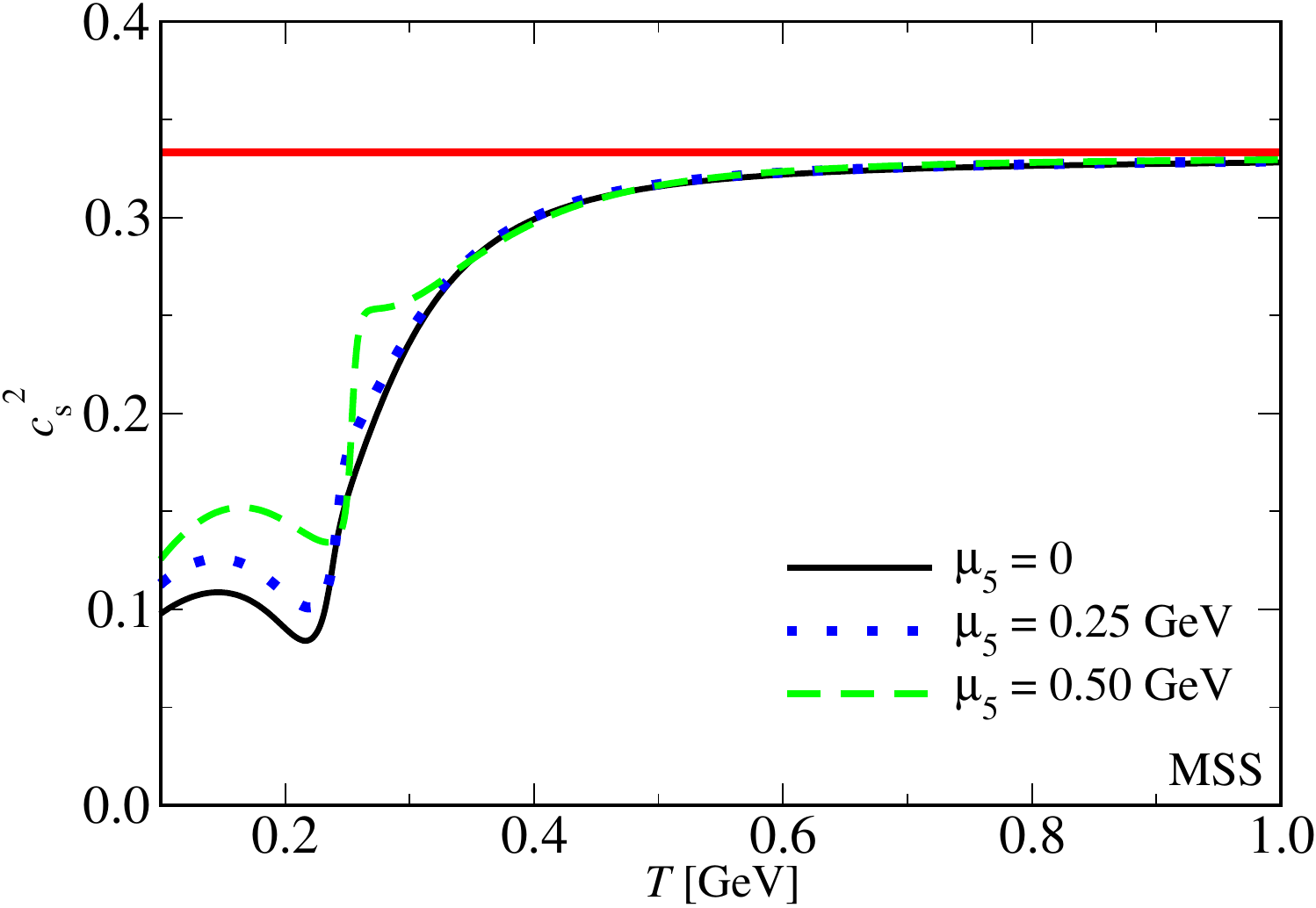}}
\caption{Scaled trace anomaly $\Delta/T^4$, equation of state
  $p_N\times \varepsilon$ and sound velocity squared as a function of
  temperature and different values of $\mu_5$ (all at $\mu=0$) in the
  TRS case, panels (a), (c) and (e), respectively and in the MSS case,
  panels (b), (d) and (f), respectively. All results obtained using
  $\mathcal{U}(\Phi,T,\mu_5)$ as given by Eq.~\eqref{Uphi2}.}
 \label{Fig8}
 \end{figure*}

Results for $p_N$, $s$, $\epsilon$, $c^2_s$, and $\Delta$  are shown in
{}Figs~\ref{Fig7} and \ref{Fig8}.  Initially, it is worth mentioning that the results for these thermodynamic quantities converge to the Stefan-Boltzmann limit, 
i.e. the limit of a gas of noninteracting particles. We recall that the Stefan-Boltzmamm limit of the pressure is given by~\cite{Costa:2009ae}
\begin{equation}
 p_{\mathrm{SB_{qg}}} =  T^4\left(p_{\mathrm{SB_{g}}}  +
 p_{\mathrm{SB_{q}}}\right),
 \label{pSB}
\end{equation}
where 
\begin{eqnarray}
 p_{\mathrm{SB_{g}}} &=& \left(N_c^2 -
 1\right)\frac{\pi^2}{45},\\ p_{\mathrm{SB_{q}}} &=& N_c N_f
 \frac{7\pi^2}{180},
\end{eqnarray}
are the contributions from the gluons and fermions respectively. The
Stefan-Boltzmann limits for the other thermodynamic quantities, which
are derived from the pressure according to the definitions~\eqref{entropy}-\eqref{cs2} given above, and are also identified in {}Figs.~\ref{Fig7}
and~\ref{Fig8}.

{}Figure~\ref{Fig7} reveals that all TRS results
show a discontinuity for $\mu_5 > 0.44$ GeV at the critical temperature,
characteristic of a first-order transition, contrary to the MSS results that predict a crossover. Although both schemes lead to results that converge to the Stefan-Boltzmann limit at high temperatures, the TRS results for the pressure, entropy, and energy density approach this limit  from above for $\mu_5 = 0.5$ GeV, for which there is a first-order transition. {}For the MSS using
$\mathcal{U}(\Phi,T,\mu_5)$,  Eq.~\eqref{Uphi2}, the curves are less
sensitive to the increase of $\mu_5$ and converge very fast, however,
they can exceed the Stefan-Boltzmann limits for some values of the
chiral chemical potential when going beyond $\mu_5 > 0.5$ GeV.

It is known that the scaled trace anomaly $\Delta$ has
peaks sharply at the deconfined temperature and has a tail that
approaches zero at asymptotically large values of $T$ in pure gauge
theories. A similar behavior is obtained for the PNJL model as shown
in {}Fig.~\ref{Fig8} for both methods, with
the caveat that at $\mu_5 = 0.5$ GeV the TRS curve diverges at
the critical temperature.  Also in {}Fig.~\ref{Fig8}, we show a
comparison of the TRS and MSS results for the equation of state (EOS) and the squared speed of sound. {}{}From Eq.~\eqref{cs2}, one can see
that $c_s^2$ represents the slope of the curve $p_n\times \varepsilon$
at each temperature value. Once again, the main difference
between the curves occurs at $\mu_5 = 0.5$ GeV for TRS, for which the speed of
sound goes to zero in the first-order transition region. To express
the pressure and energy density for the same temperature range used
for $c_s^2$ it is more convenient to use a logarithmic scale in
the EOS plot. This is the reason why the jump at $p_N\lesssim $ 0.1
GeV/fm$^3$ appears to be more pronounced than the equivalent discontinuity in
the squared speed of sound. {}On the other hand, the MSS curve does not
show a first-order transition; the steplike behavior in the region
$0.2~{\rm GeV} \lesssim T \lesssim 0.3~{\rm GeV}$ is related to a continuous change
of slope of the EOS in the range $0.8~{\rm GeV/fm}^3 \lesssim \varepsilon \lesssim
10~{\rm GeV/fm}^3$.  {}For the other values of $\mu_5$ in both methods,
the curves approach the conformal limit of 1/3 at values of $T$ of the
order of $0.6$ GeV. 

\begin{center} 
\begin{figure}[!htpb]
 \subfigure[]{\includegraphics[scale=0.33]{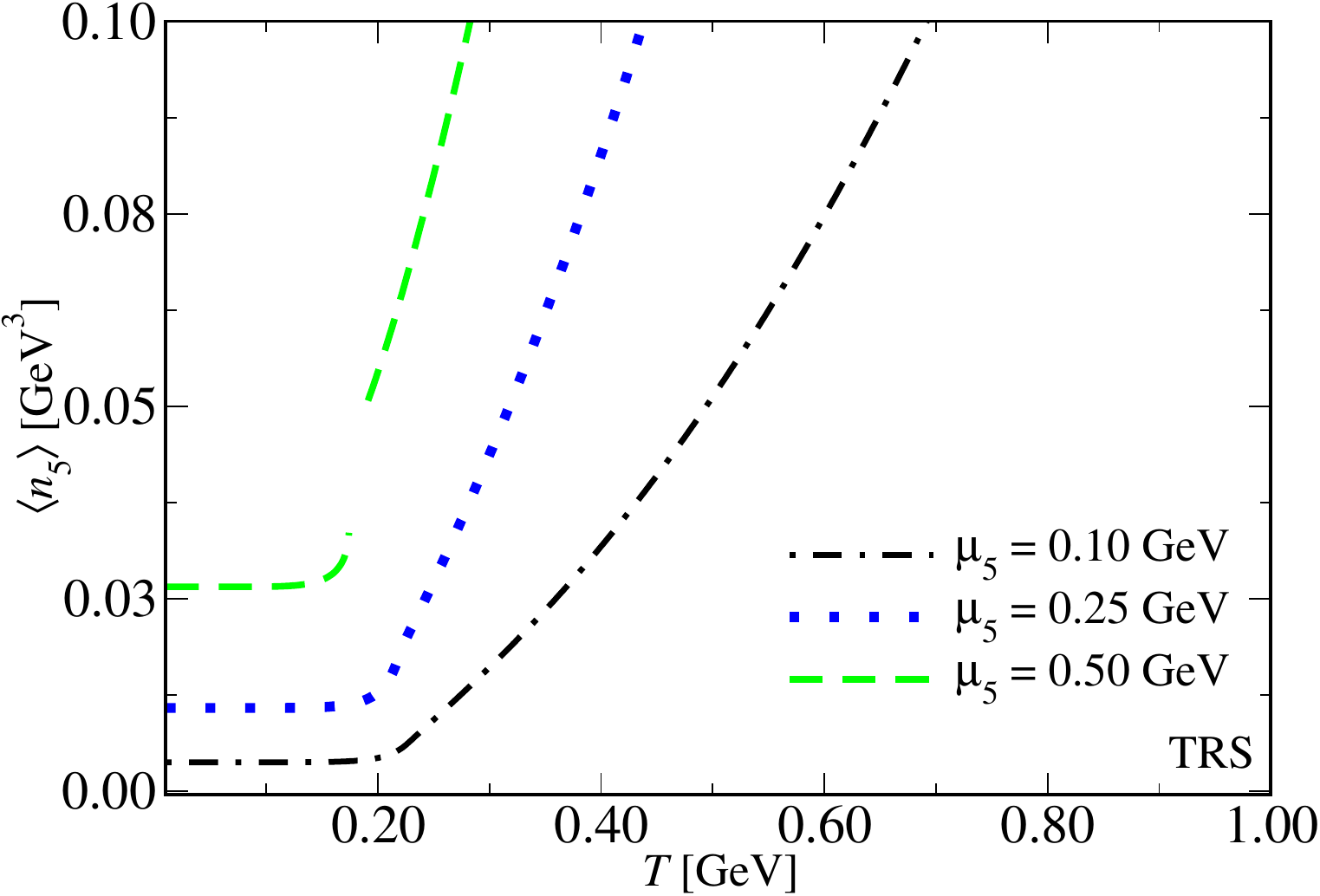}}
 \subfigure[]{\includegraphics[scale=0.33]{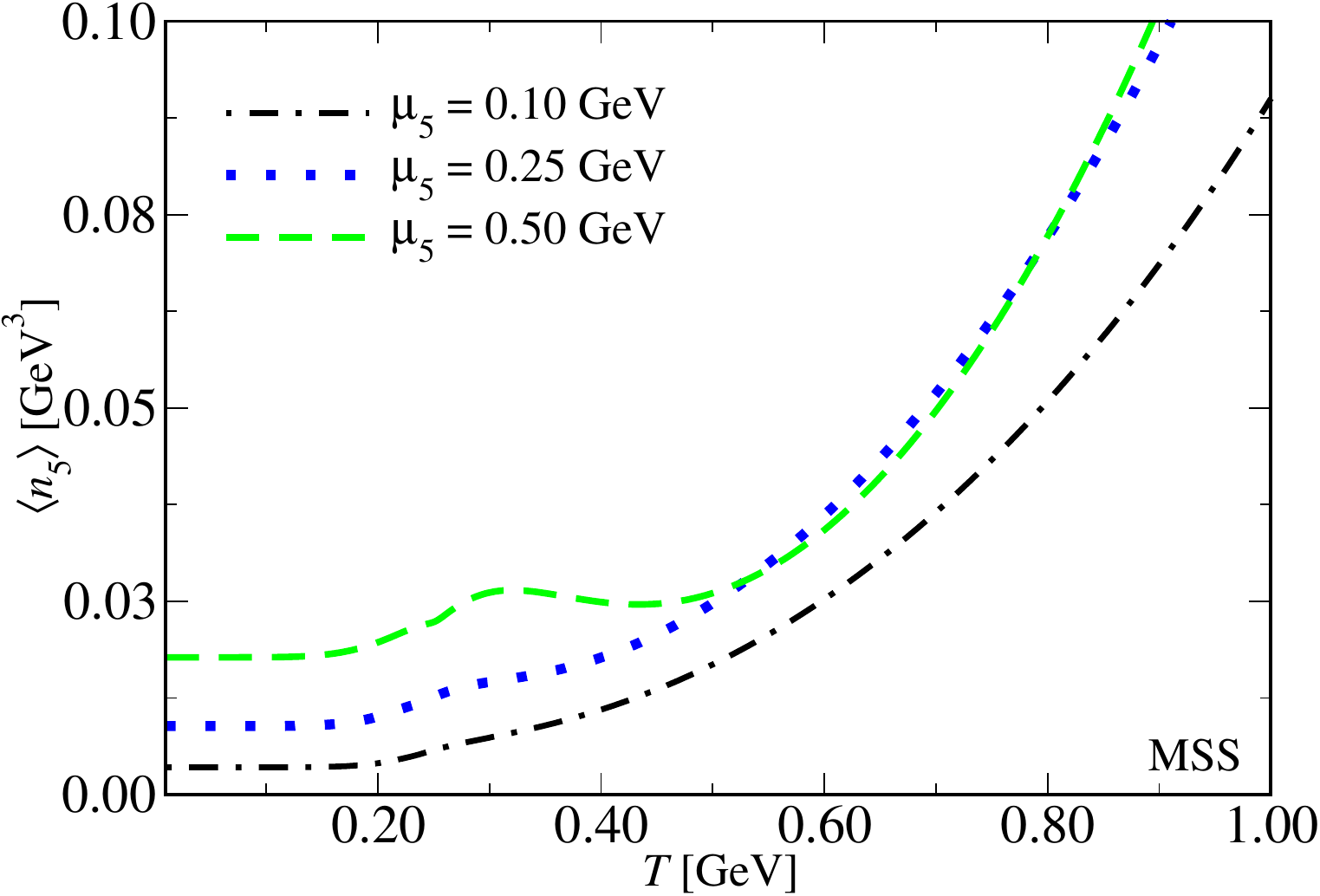}}
 \caption{Chiral density from Eq.~\eqref{n5full} as a function of the
   temperature $T$, comparing TRS, on panel (a) and MSS, on panel (b),
   for different values of $\mu_5$.}
 \label{Fig9}
 \end{figure} 
\end{center} 

In {}Fig.~\ref{Fig9} we show the results for the chiral density $\langle n_{5}\rangle$,
defined in Eq.~\eqref{n5full}, in both the TRS and MSS methods and using the redefined Polyakov loop
given by Eq.~\eqref{Uphi2}.  This is the quantity for which the
differences between TRS and MSS are more pronounced. In the TRS case,
for all the values of $\mu_5$ considered, the curves have a similar behavior as a function of temperature: a constant line with a change in concavity at the
transition temperature, succeeded by a monotonic increase. {}For $\mu_5 = 0.5$ GeV this quantity also presents a
small jump at $T_{\rm pc}$, characteristic of a first-order
transition.  In the MSS case, $\langle n_{5}\rangle$ is an
increasing function of the temperature for small values of $\mu_5$,
and shows a small hump in the pseudocritical temperature region for
higher values of of $\mu_5$. We do not show the results for $\mu_5 = 0$ because $\langle n_{5}\rangle $ is trivially zero in this case.

\section{Phase diagram at finite quark chemical potential}
\label{sec4}

{}Finally, we consider the influence on the phase diagram in the 
$T\times\mu$ plane of including the dependence on $\mu_5$
in the Polyakov-loop potential as indicated in Eq.~(\ref{b2mu5}). In this case, we have $\Phi\neq\Phi^{\dagger}$ and, instead of~Eq.~\eqref{Tpc}, we need to solve
\begin{equation}
\left.\frac{\partial^2M}{\partial T^2}\right|_{T =
  T^{c}_{\rm pc}} = \left.\frac{\partial^2\Phi}{\partial T^2}\right|_{T =
  T^{\Phi}_{\rm pc}} = \left.\frac{\partial^2\Phi^{\dagger}}{\partial
  T^2}\right|_{T = T^{\Phi^{\dagger}}_{\rm pc}} = 0 .
\end{equation}
We define the deconfinement pseudocritical temperature $T_{\rm pc}^d$ as 
the average of $T^{\Phi}_{\rm pc}$ and $T^{\Phi^\dagger}_{\rm pc}$:
\begin{equation}
T_{\rm pc}^d = \frac{T^{\Phi}_{\rm pc} + T^{\Phi^{\dagger}}_{\rm
    pc}}{2}\,.
\end{equation}

\begin{center} 
\begin{figure}[!htbp]
\begin{center}
\subfigure[]{\includegraphics[scale=0.4]{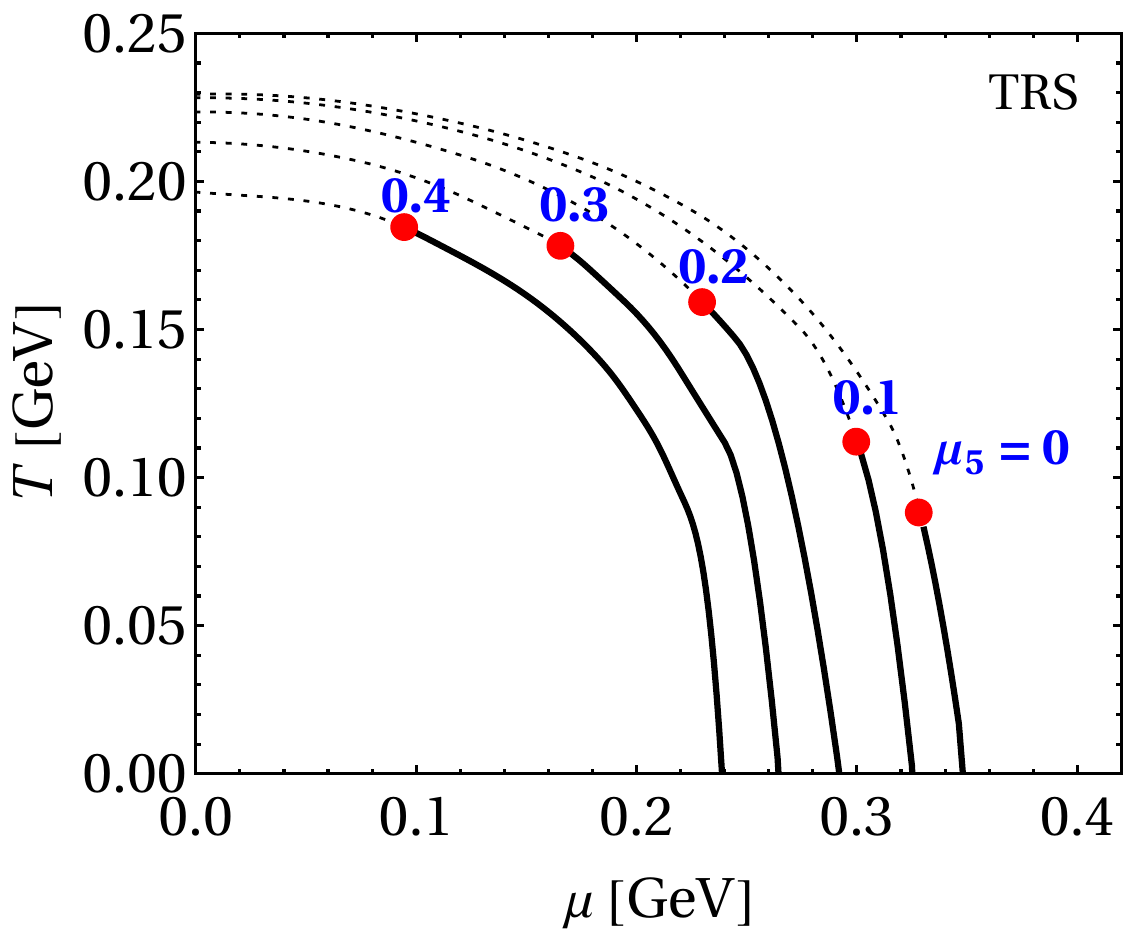}}
\subfigure[]{\includegraphics[scale=0.4]{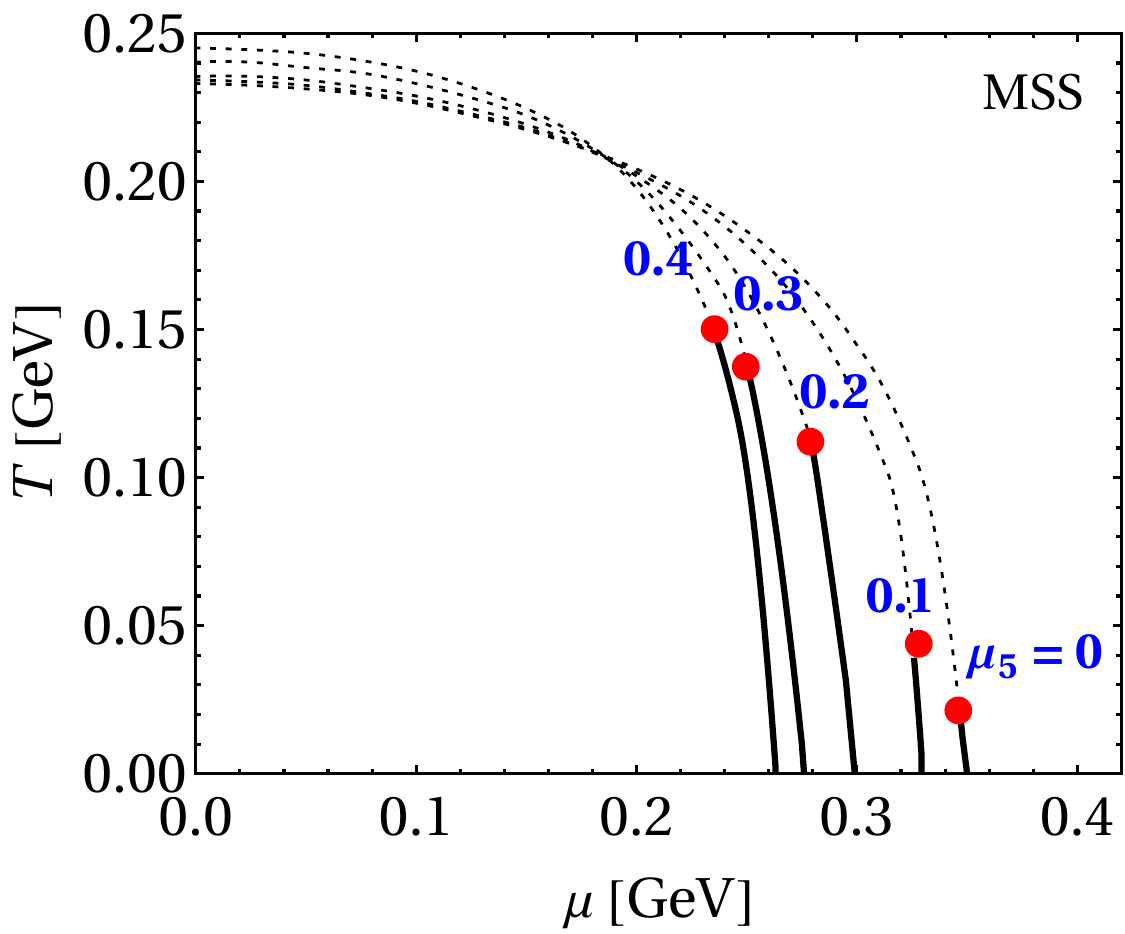}}
\caption{The evolution of the CEP in the space $T\times\mu$, for
  different values of $\mu_5$, comparing TRS, on panel (a), and MSS,
  on panel (b). Dashed and solid lines correspond to second and
  first-order transitions, respectively, while the large dot is the
  CEP. The values of $\mu_5$ are presented right above the CEPs, and
  their values are given in GeV.}
\label{Fig10}
\end{center}
\end{figure}
\end{center} 

{}Figure~\ref{Fig10} shows the evolution of the CEP for different
values of $\mu_5$. One sees that the
CEP is shifted toward lower values of $\mu$ and higher values of $T$ when
$\mu_5$ increases. The same behavior is observed for TRS using
$\mathcal{U}(\Phi,\Phi^{\dagger},T)$ from Eq.~\eqref{Uphi} and for
MSS using $\mathcal{U}(\Phi,\Phi^{\dagger},T,\mu_5)$
from Eq.~\eqref{Uphi2}, although the positions of the critical end points and the
slopes of the curves are different for each scheme.

\section{Conclusions}
\label{conclusions}

In this paper, we performed a thorough study within the PNJL model on the differences
that the TRS and MSS regularization procedures imply for the thermodynamics of quark matter with chiral imbalance.  We have shown that, to reproduce lattice results for the chiral and deconfinement pseudocritical temperatures as a function of the chiral chemical
potential~$\mu_5$, it is necessary to include a $\mu_5$-dependece in the parametrization of the quadratic $\Phi^\dagger\Phi$ term in Polyakov-loop potential, in addition to the usual temperature dependence. We have proposed a polynomial dependence analogous to that in terms of temperature. In addition to reproducing the lattice results for values of the pseudocritical temperatures, the proposed parametrization also reproduces the increase of those temperatures with $\mu_5$, indicating that such a $\mu_5$ dependence in the Polyakov-loop potential may be the missing ingredient in the traditional PNJL models. We have studied the changes caused by this new parametrization of the Polyakov-loop potential in various thermodynamic quantities, and also contrasted the results predicted by the TRS and MSS regularization procedures. In particular, we have shown that the combined use of the new parametrization of the Polyakov-loop  potential and the MSS procedure leads to consistent results that agree with those of lattice simulations.
\section*{Acknowledgements}

This work was partially supported by Conselho Nacional de
Desenvolvimento Cient\'{\i}fico e Tecnol\'ogico (CNPq), Grants
No. 312032/2023-4 (R.L.S.F.), No. 307286/2021-5 (R.O.R.) and No. 309262/2019-4 (G.K.); Funda\c{c}\~ao
de Amparo \`a Pesquisa do Estado do Rio Grande do Sul (FAPERGS),
Grants No. 19/2551-0000690-0 and 19/2551-0001948-3 (R.L.S.F.), and
23/2551-0000791-6 and 23/2551-0001591-9 (D.C.D.); Funda\c{c}\~ao de Amparo \`a Pesquisa do Estado de S\~ao Paulo (FAPESP) Grant No. 2018/25225-9 (G.K.); Funda\c{c}\~ao
Carlos Chagas Filho de Amparo \`a Pesquisa do Estado do Rio de Janeiro
(FAPERJ), Grant No. E-26/201.150/2021 (R.O.R.); Coordena\c{c}\~ao de
Aperfei\c{c}oamento de Pessoal de N\'ivel Superior - Brasil (CAPES) -
Finance Code 001 (F.X.A.). The work is also part of the project
Instituto Nacional de Ci\^encia e Tecnologia - F\'isica Nuclear e
Aplica\c{c}\~oes (INCT - FNA), Grant No. 464898/2014-5 and supported
by the Ser\-ra\-pi\-lhei\-ra Institute (Grant No. Serra -
2211-42230).



\end{document}